\documentclass[twocolumn]{aastex63}
\usepackage{newtxtext,newtxmath}
\usepackage[T1]{fontenc}
\usepackage{CJKutf8}
\usepackage{ae,aecompl}
\usepackage{graphicx}      
\usepackage{amsmath}
\usepackage{amssymb}	
\usepackage{float}
\usepackage[]{hyperref}

\usepackage{booktabs}
\received{}
\revised{}
\accepted{}

\submitjournal{ApJ} 
\begin{document}

\title{The $EB$-correlation in Resolved Polarized Images: Connections to Astrophysics of Black Holes}

\date{May 2023}

\shorttitle{EB-correlation}
\shortauthors{R. Emami et. al.}

\correspondingauthor{Razieh Emami}
\email{razieh.emami$_{-}$meibody@cfa.harvard.edu}

\author[0000-0002-2791-5011]{Razieh Emami}
\affiliation{Center for Astrophysics $\vert$ Harvard \& Smithsonian, 60 Garden Street, Cambridge, MA 02138, USA}

\author[0000-0002-9031-0904]{Sheperd S. Doeleman}
\affiliation{Center for Astrophysics $\vert$ Harvard \& Smithsonian, 60 Garden Street, Cambridge, MA 02138, USA}
\affiliation{Black Hole Initiative at Harvard University, 20 Garden Street, Cambridge, MA 02138, USA}

\author[0000-0002-8635-4242]{Maciek Wielgus}
\affiliation{Max-Planck-Institut f\"ur Radioastronomie, Auf dem H\"ugel 69, D-53121 Bonn, Germany}

\author[0000-0001-9939-5257]{Dominic Chang}
\affiliation{Center for Astrophysics $\vert$ Harvard \& Smithsonian, 60 Garden Street, Cambridge, MA 02138, USA}
\affiliation{Black Hole Initiative at Harvard University, 20 Garden Street, Cambridge, MA 02138, USA}

\author[0000-0002-2825-3590]{Koushik Chatterjee}
\affiliation{Black Hole Initiative at Harvard University, 20 Garden Street, Cambridge, MA 02138, USA}

\author[0000-0003-4284-4167]{Randall \ Smith}
\affiliation{Center for Astrophysics $\vert$ Harvard \& Smithsonian, 60 Garden Street, Cambridge, MA 02138, USA}

\author[0000-0003-4475-9345]{Matthew Liska}
\affiliation{Center for Astrophysics $\vert$ Harvard \& Smithsonian, 60 Garden Street, Cambridge, MA 02138, USA}

\author[0000-0002-5872-6061]{James F. Steiner}
\affiliation{Center for Astrophysics $\vert$ Harvard \& Smithsonian, 60 Garden Street, Cambridge, MA 02138, USA}

\author[0000-0001-5287-0452]{Angelo Ricarte}
\affiliation{Black Hole Initiative at Harvard University, 20 Garden Street, Cambridge, MA 02138, USA}
\affiliation{Center for Astrophysics $\vert$ Harvard \& Smithsonian, 60 Garden Street,  Cambridge, MA 02138, USA}

\author[0000-0002-1919-2730]{Ramesh Narayan}
\affiliation{Center for Astrophysics $\vert$ Harvard \& Smithsonian, 60 Garden Street, Cambridge, MA 02138, USA}
\affiliation{Black Hole Initiative at Harvard University, 20 Garden Street, Cambridge, MA 02138, USA}

\author[0000-0001-6950-1629]{Grant Tremblay}
\affiliation{Center for Astrophysics $\vert$ Harvard \& Smithsonian, 60 Garden Street,  Cambridge, MA 02138, USA}

\author[0000-0003-2808-275X]{Douglas Finkbeiner}
\affiliation{Center for Astrophysics $\vert$ Harvard \& Smithsonian, 60 Garden Street, Cambridge, MA 02138, USA}

\author[0000-0001-6950-1629]{Lars \ Hernquist}
\affiliation{Center for Astrophysics $\vert$ Harvard \& Smithsonian, 60 Garden Street,  Cambridge, MA 02138, USA}

\author[0000-0001-6337-6126]{Chi-Kwan Chan}
\affiliation{Steward Observatory and Department of Astronomy, University of Arizona, 933 N. Cherry Ave., Tucson, AZ 85721, USA}
\affiliation{Data Science Institute, University of Arizona, 1230 N. Cherry Ave., Tucson, AZ 85721, USA}
\affiliation{Program in Applied Mathematics, University of Arizona, 617 N. Santa Rita, Tucson, AZ 85721}

\author[0000-0002-9030-642X]{Lindy Blackburn}
\affiliation{Center for Astrophysics $\vert$ Harvard \& Smithsonian, 60 Garden Street, Cambridge, MA 02138, USA}
\affiliation{Black Hole Initiative at Harvard University, 20 Garden Street, Cambridge, MA 02138, USA}

\author[0000-0002-0393-7734]{Ben S. Prather}
\affiliation{CCS-2, Los Alamos National Laboratory, P.O. Box 1663, Los Alamos, NM 87545, USA}

\author[0000-0003-3826-5648]{Paul Tiede}
\affiliation{Center for Astrophysics $\vert$ Harvard \& Smithsonian, 60 Garden Street, Cambridge, MA 02138, USA}
\affiliation{Black Hole Initiative at Harvard University, 20 Garden Street, Cambridge, MA 02138, USA}

\author[0000-0002-3351-760X]{Avery E. Broderick}
\affiliation{Perimeter Institute for Theoretical Physics, 31 Caroline Street North, Waterloo, ON, N2L 2Y5, Canada}
\affiliation{Department of Physics and Astronomy, University of Waterloo, 200 University Avenue West, Waterloo, ON, N2L 3G1, Canada}

\author[0000-0001-8593-7692]{Mark Vogelsberger}
\affiliation{Department of Physics, Kavli Institute for Astrophysics and Space Research, Massachusetts Institute of Technology, Cambridge, MA 02139, USA}

\author[0000-0002-7892-3636]{Charles Alcock}
\affiliation{Center for Astrophysics $\vert$ Harvard \& Smithsonian, 60 Garden Street, Cambridge, MA 02138, USA}

\author[0000-0001-5461-3687]{Freek Roelofs}
\affiliation{Black Hole Initiative at Harvard University, 20 Garden Street, Cambridge, MA 02138, USA}
\affiliation{Center for Astrophysics $\vert$ Harvard \& Smithsonian, 60 Garden Street,  Cambridge, MA 02138, USA}

\begin{abstract}
We present an in-depth analysis of a newly proposed correlation function in visibility space, between the $E$ and $B$ modes of the linear polarization, hereafter the $EB$-correlation, for a set of time-averaged GRMHD simulations compared with the phase map from different semi-analytic models as well as the Event Horizon Telescope (EHT) 2017 data for M87* source. We demonstrate that the phase map of the time-averaged $EB$-correlation contains novel information that might be linked to the BH spin, accretion state and the electron temperature. A detailed comparison with a semi-analytic approach with different azimuthal expansion modes shows that to recover the morphology of the real/imaginary part of the correlation function and its phase, we require higher orders of these azimuthal modes. To extract the phase features, we propose to use the Zernike polynomial reconstruction developing an empirical metric to break degeneracies between models with different BH spins that are qualitatively similar. We use a set of different geometrical ring models with various magnetic and velocity field morphologies and show that both the image space and visibility 
based $EB$-correlation morphologies in MAD simulations can be explained with simple fluid and magnetic field geometries as used in ring models. SANEs by contrast are harder to model, demonstrating that the simple fluid and magnetic field geometries of ring models are not sufficient to describe them owing to higher Faraday Rotation depths. A qualitative comparison with the EHT data demonstrates that some of the features in the phase of $EB$-correlation might be well explained by the current models for BH spins as well as electron temperatures, while others may require a larger theoretical surveys.

\end{abstract}


\section{Introduction}
The recently published image of the giant supermassive black hole (SMBH) at the center of the elliptical galaxy Messier 87 \citep[hereafter M87*;][]{PaperI, PaperII, PaperIII, PaperIV, PaperV, PaperVI} by the Event Horizon Telescope Collaboration (EHTC) reveals an asymmetric ring-like structure with a diameter of 42 $\pm$ 3 $\mu$as, consistent with the predicted shadow size of a BH from the Einstein theory of general relativity (GR) \citep{PaperV, PaperVI}. The ring-like structure appears to be a persistent image feature on timescales of years, further corroborating this interpretation \citep{Wielgus2020}. The EHT image is generally consistent with the common picture in which the environment of M87* is made of a hot, relativistic, and magnetized plasma \citep{1982Natur.295...17R, Narayan+1995,1995Natur.374..623N}. However, from the intensity map, it is not possible to constrain the morphology of the magnetic field nor identify whether the compact emission might have originated from inflowing matter or rather from an outflow, corresponding to thick accretion or jet/wind-based emission, respectively.  

The spatially resolved linear polarization pattern reported by the \cite{PaperVII, PaperVIII} 
provides very useful information about the orientation of the electric vector polarization angle (EVPA). Furthermore, it also reveals an (almost) azimuthally symmetric pattern for EVPAs. This motivated \cite{Palumbo_2020} to propose a particular decomposition for the linear polarization in terms of the azimuthal modes, as specified with complex coefficient $\beta_m$. They showed that $m=2$ provides the dominant contribution in the EVPAs. Moreover, they found  that the amplitude and phase of the $\beta_2$ coefficient are linked to the magnetic field geometry and BH spin. However, they did not characterize the key driver of $\beta_2$. \cite{Emami2022} made a comprehensive study of the origin of the twisty patterns in linear polarization using a set of spatially resolved (time-averaged) images of General Relativistic Magneto-Hydro Dynamical (GRMHD) simulations and identified the magnetic field geometry as the key contributor to creating the twisty patterns in linear polarization, as approximated with $\beta_2$. While $\beta_2$ well-characterizes the salient rotationally-symmetric features of the linearly polarized structure, it neglects other asymmetric features of the image.

In this work, we propose a new correlation function, the $EB$-correlation in visibility space for the EHT 2017 data as well as the time-averaged images of GRMHD simulations. In the latter case, we use different approaches including both a full-numerical algorithm as well as a semi-analytic method, comparing the real and the imaginary parts of the correlation function using different cut-offs in the azimuthal expansion of the $E$ and $B$ modes. We find that higher order terms are indeed required in the azimuthal expansion of $E$ and $B$  modes. We use a Zernike reconstruction algorithm to recover the $EB$-correlation phase map and propose an empirical metric to  extract features from the phase map. Our analysis demonstrates that models that are qualitatively similar exhibit different patterns which may be eventually useful for recovering the BH spin. We use a family of different geometrical ring models, each with different $B$ and $V$ field morphologies, and identify a case that almost captures the features in both the image and visibility spaces. We make a qualitative comparison between the $EB$-correlation function from the EHT 2017 data as well as the time-averaged images of GRMHD simulations and show some of the features in the phase map can be well described by the limited library of images, as considered in this study, while others may require additional explorations as is left for some future studies. 

The paper is structured as follows. Section \ref{E-B-terminology} presents the key steps in defining the $EB$-correlation in the visibility space. In Section \ref{EB-real-data} we show an immediate application to the EHT 2017 data. 
Section \ref{Toy-Model-Set} constructs the $EB$-correlation phase for a set of few toy models.
In Section \ref{Numerics} we present our numerical tools for making the polarized BH images. Section \ref{ZP-expansion} introduces the Zernike polynomials to extract the features from the correlation phase. In Section \ref{Correlation-Phase} we infer the correlation phase for the time-averaged images. Section \ref{morder-expansion} presents a semi-analytic model based on m-order expansions of the $E$ and $B$ modes and compares them with the full-numerical analysis. In Section \ref{Geometrical-Ring} we present a geometrical ring model that best captures both the time-averaged images and the $EB$-correlation phase. Section \ref{Com-EHT-data} provides a comparison with the EHT 2017 data. In Section \ref{conclusion} we present the conclusion of the paper. Section \ref{future} discusses future directions for this new correlation function. We present some technical details in Appendices \ref{EHT-resolution}-\ref{phase-semi-analysis}. 

\section{$EB$-correlation function in the visibility space}
\label{E-B-terminology}
In the following, we define the $E$ and $B$ modes and construct their cross-correlation function in visibility space. 
\subsection{Definition: $E$ mode and $B$ mode}
As the first step, here we construct the  flat sky based
$E$ and $B$ modes in visibility space. We follow the notation of \cite{2016ARA&A..54..227K, PaperVIII} in which $E$ and $B$ modes are defined naturally in the visibility space $(u,v)$, where $u$ and $v$ are sampled based on the interferometer vector. In this representation, the $E$ and $B$ visibilities (note that throughout our analysis we have dropped the tilde from $E$ and $B$ modes as they are naturally defined in the visibility space) are related to the Stokes parameters $\tilde{\mathcal{Q}}(u,v)$ and $\tilde{\mathcal{U}}(u,v)$ by a local rotation of $2\theta$ in the Fourier space:
\begin{equation}
\label{EB-QU}
\begin{pmatrix} E(u, \theta)\\\ B(u, \theta)\end{pmatrix}=\begin{pmatrix}\cos2\theta&\sin2\theta\\-\sin2\theta&\cos2\theta\end{pmatrix}\begin{pmatrix}\tilde{\mathcal{Q}}(u, \theta)\\\tilde{\mathcal{U}}(u, \theta)\end{pmatrix},
\end{equation}
where the rotation angle is defined as \footnote{In this paper, we use the convention taken in the ehtim  package \citep{2018ApJ...857...23C}.}:
\begin{equation}
\label{rotation-angle}
\theta \equiv \arctan{\left(\frac{u}{v}\right)} .
\end{equation}
In addition, in our analysis, we use the relation between the real and the Fourier space quantities, 
\begin{equation}
\label{real-Fourier}
\tilde{X}(u, \theta) = \int\int X(\rho, \phi) e^{-2i\pi \rho u \cos{(\theta - \phi)}} \rho d\rho d\phi ,
\end{equation}
where $X$ is a generic function in the image plane, e.g. the Stokes parameter.\footnote{Notice that in the image plane, we have defined $\phi$ from $y$ axis rather than $x$.
}.

While for our analysis we merely focus on the $E$ and $B$ modes in the visibility space, it is important to also comment on their definitions in real space;
where the $E$ and $B$ modes refer to the gradient and  curl components of the polarization field, respectively, 
\cite[see Eqs. (A2) and (A3) of][for more details]{PaperVIII}. Quite importantly, while the real space $Q$ and $U$ depend on the choice of the coordinate system, the real space $E$ and $B$ modes are scalar quantities; i.e. invariant under the change of the coordinate system.

 Below, we make an in-depth analysis of
constructing the $EB$-correlation function using different approaches; aiming to extract information about plasma astrophysics. 

\subsection{Construction: $EB$-correlation function}
We define the normalized $EB$-correlation function in visibility space as: 
\begin{equation}
\label{EB-correlation}
\rho_{\mathcal{EB}}(u,v) \equiv \frac{E(u, v) B^{*}(u, v)}{\sqrt{E(u,v)  E^*(u,v)} \sqrt{B(u,v) B^*(u,v)}} .
\end{equation}
As this is a complex quantity, we infer its phase from:\footnote{To capture the sign of the real and the imaginary part of the $EB$-correlation, in practice, we use the $\mathrm{arctan2(imaginary, real)}$ function from the python numpy library.} 
\begin{equation}
\label{EB-phase}
\theta(u,v) \equiv \arctan{\left( \frac{Im(\rho_{\mathcal{EB}}(u,v))}{Re(\rho_{\mathcal{EB}}(u,v))} \right)}.
\end{equation} 
While Eq. (\ref{EB-phase}) specifies the correlation phase for a given function $\rho_{\mathcal{EB}}(u,v)$, it does not immediately provide an algorithm for how to compute this function for the time-averaged images of GRMHD simulations. This is important as computing the phase average may be ill-defined due to phase wrapping \citep{PaperVII}. To avoid this issue, below we introduce two different metrics for calculating the correlation phase from the time-averaged images of GRMHD simulations. 
\begin{figure*}[th!]
\center
\includegraphics[width=0.89\textwidth]{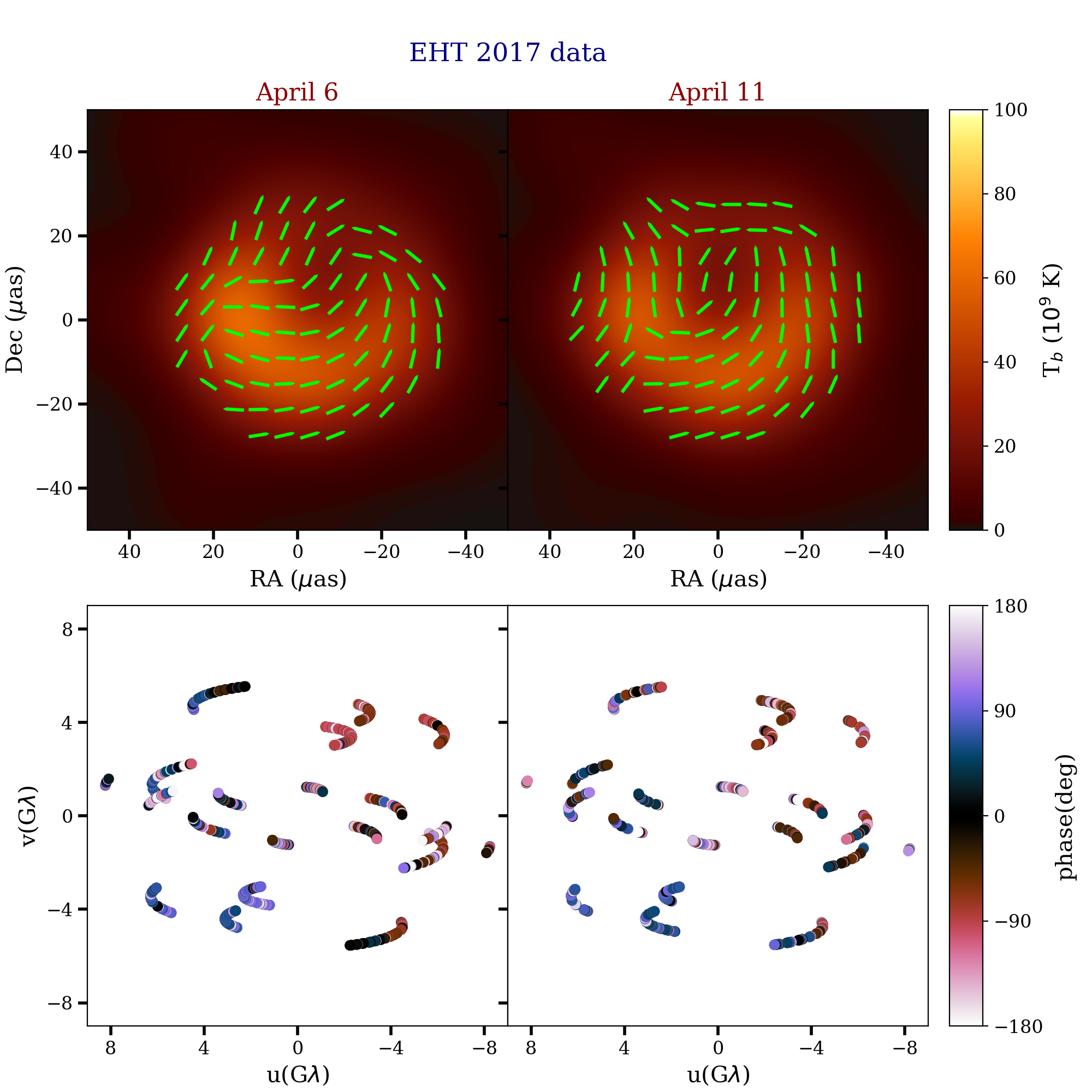}
\caption{ The polarized image of M87* from EHT 2017 data on April 6 and April 11 as well as the $EB$-correlation phase in the visibility space (bottom). }
\label{EHT-data}
\end{figure*}
\subsubsection{Time-averaged Correlation phase: First method} \label{phase-1st-method}
In this method, we first compute the time-averaged $E$ and $B$ modes from the GRMHD simulations. We subsequently infer the correlation function using Eq. (\ref{EB-phase}). The phase is therefore given by:
\begin{equation}
\label{mean-EB-phase-1st}
\bar{\theta}_{\mathrm{1st}}(u,v) = \arctan{\left( \frac{  Im(\rho_{\mathcal{\langle E \rangle \langle B \rangle}}(u,v))}{ Re(\rho_{\mathcal{\langle E \rangle \langle B \rangle}}(u,v))} \right)},
\end{equation} 
where in Eq. (\ref{mean-EB-phase-1st}), $\mathcal{\langle E \rangle}$ and $\mathcal{\langle B \rangle}$ refer to the time-averaged $E$ and $B$ modes, respectively.
 Throughout this paper, we use this method as the main approach for computing the $EB$-correlation phase. Hereafter we call this the $M_1$ method. 

\subsubsection{Time-averaged Correlation phase: Second method}
\label{phase-2nd-method}
In this approach, we first calculate the time-averaged correlation function and subsequently infer the correlation phase using Eq. (\ref{EB-phase}):
\begin{equation}
\label{mean-EB-phase-2nd}
\bar{\theta}_{\mathrm{2nd}}(u,v) = \arctan{\left( \frac{ \langle Im(\rho_{\mathcal{EB}}(u,v)) \rangle}{ \langle Re(\rho_{\mathcal{EB}}(u,v)) \rangle } \right)},
\end{equation} 
where in Eq. (\ref{mean-EB-phase-2nd}), $\langle Im(\rho_{\mathcal{EB}}(u,v)) \rangle$ refers to the time-averaged correlation function. 

As stated above, in what follows we mainly focus on the first approach. However, in Section \ref{second-approach-phase} we infer the phase adopting the second approach to make a comparison and justify that differences are not substantial. Hereafter we call this the $M_2$ method. 

\section{$EB$-correlation for EHT 2017 data}
\label{EB-real-data}
In this section, we take the first look at the phase of the $EB$-correlation function using the fully polarized EHT 2017 data for M87*. We focus on the observations performed on April 6 and April 11 in 2017. The data set was first calibrated using the \texttt{EHT-HOPS} data pipeline \citep{Blackburn2019, PaperIII}, with additional polarization and leakage calibration as detailed in \citet{PaperVII, Issaoun2022, Jorstad2023}. There were minor updates to the EHT calibration pipeline with respect to \citet{PaperI, PaperVII}, as described in \citet{EHTC_SgrA_Paper2}.

Figure \ref{EHT-data} presents the intensity map with the EVPA ticks in the image plane (top row) as well as the $EB$-correlation phase in the visibility space (bottom row), respectively. Distinct features are visible in the discrete phase maps.  These features appear generally consistent between April 6 and April 11. Motivated by this, in the remainder of this paper, we make synthetic data, using the time-averaged images of the GRMHD simulations of H-AMR using the EHT coverage from April 11 and we perform an in-depth analysis of the $EB$-correlation function, comparing results with those derived from EHT 2017 data. 

\begin{figure*}[th!]
\center
\includegraphics[width=\textwidth]{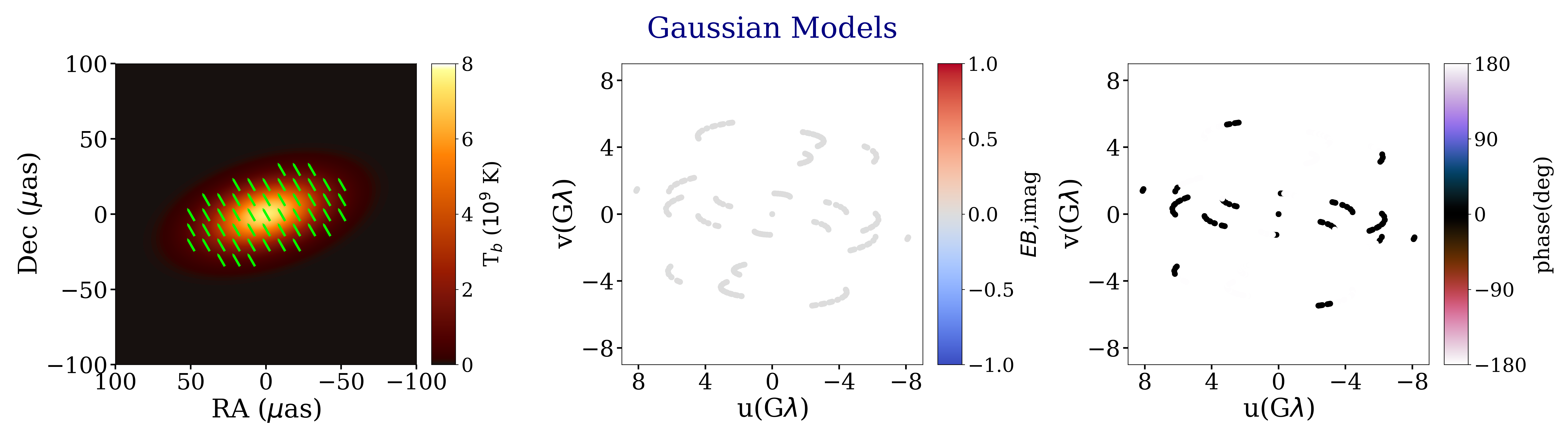}
\includegraphics[width=\textwidth]{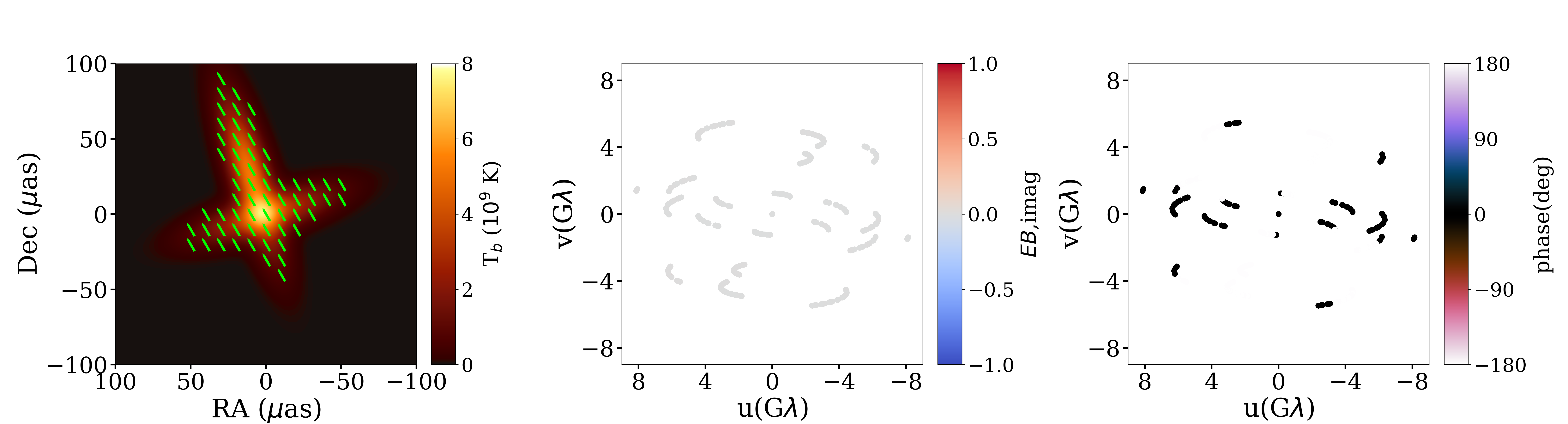}
\includegraphics[width=\textwidth]{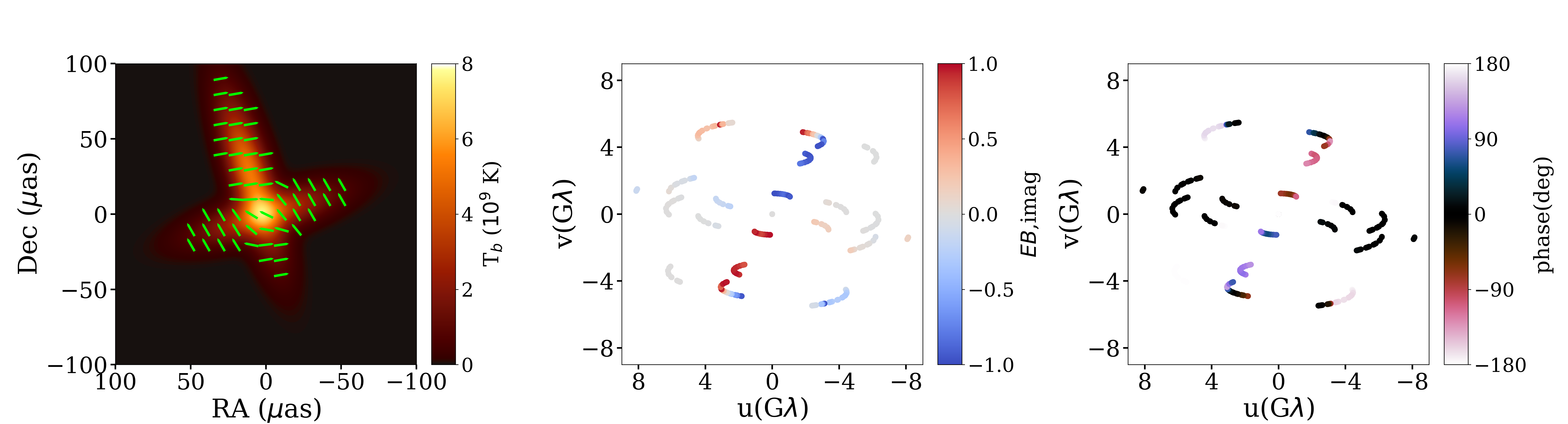}
\caption{A family of different Gaussian models. 
For a single Gaussian model, top panel, the phase of the $EB$-correlation is either zero or 180 deg. In a double Gaussian model, middle panel, with the same EVPA ticks, the phase remains the same despite the brightness asymmetry. Finally, for a double Gaussian model with different EVPA ticks, the bottom panel, the phase gets non-vanished owing to the polarization anisotropies. }
\label{Gaussian_Model}
\end{figure*}

\section{$EB$-correlation: construction of toy models}
\label{Toy-Model-Set}
As the first theory step, here we study the $EB$-correlation function for a set of different geometrical toy models, including different Gaussian models as well as a subset of geometrical ring models. 

\subsection{Gaussian models}
\label{Gaussian}
To get an intuition on what may drive non-zero correlation phase, here we study a set of different Gaussian models. Our study covers a single Gaussian (top row), a double Gaussian with the same ticks of the EVPA (middle row) and a double Gaussian model with two different sets of the EVPA ticks (down row) in Figure \ref{Gaussian_Model}, respectively. It is seen that the phase of the $EB$-correlation function remains zero or 180 deg for the first two cases while it deviates from these values for the last case. This illustrates that the asymmetries in the EVPA ticks, as against the asymmetries in the intensity, drive non-zero $EB$-correlation phases for Gaussian models. 

\begin{figure*}[th!]
\center
\includegraphics[width=\textwidth]{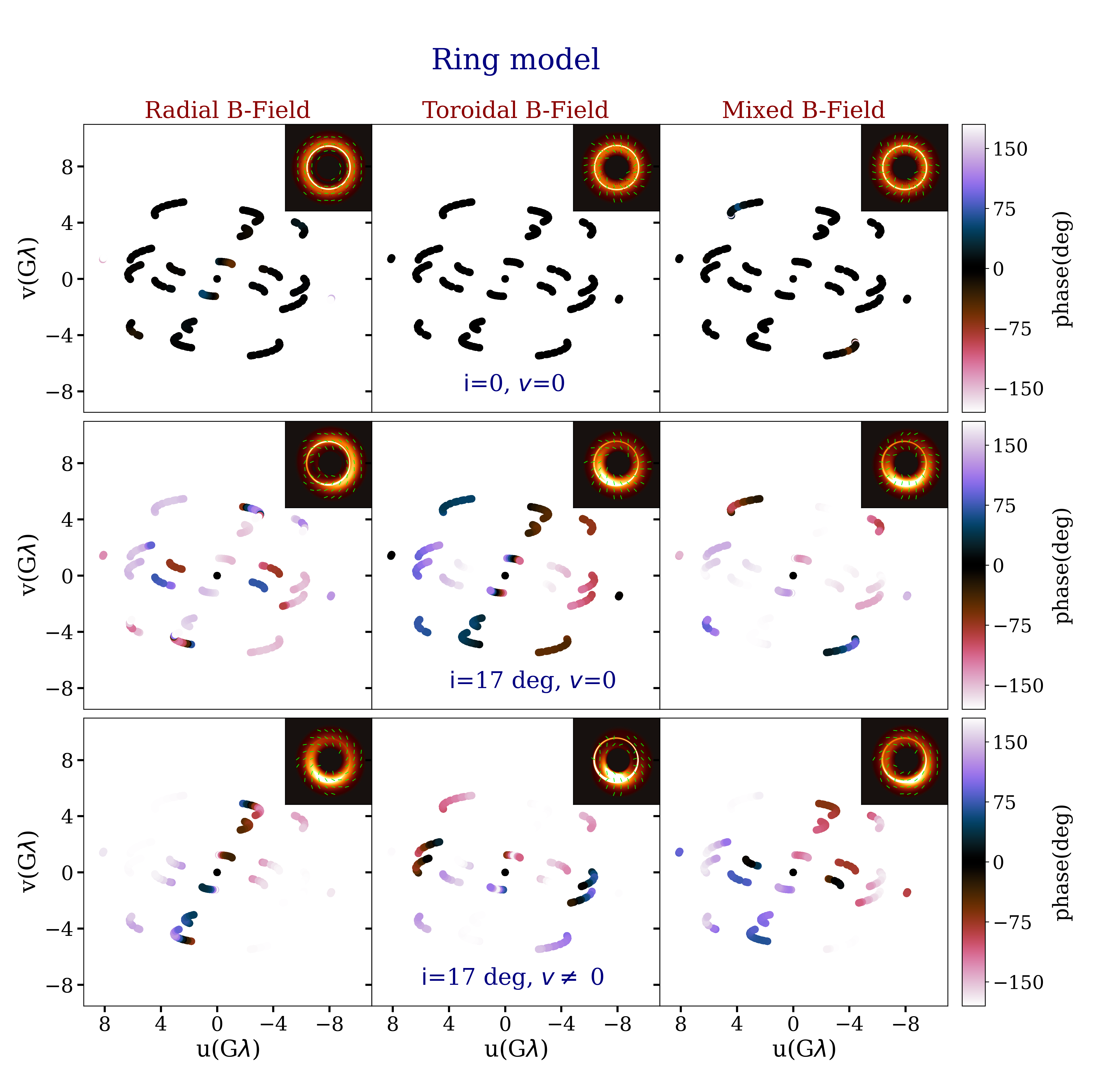}
\caption{The $EB$-correlation phase for models with zero incliation and fluid speed (top row), models with $\mathrm{i}$=17 deg while no fluid speed (middle row) and models with $\mathrm{i}$=17 deg and non-zero fluid speed bottom row, respectively. I each row, from the left to right we study different magnetic field morphologies including a radial, toroidal and a mixed magnetic field, respectively. In each panel we overlay the ring image, on the top right corer, to build an intuition about the geometry of the ring. The phase is zero for cases with zero inclination, while it is non-zero for models with non-zero inclination. It is also degenerate between models with different B and v field geometries. }
\label{G-Ring_Model}
\end{figure*}

\subsection{Geometrical Ring models}
\label{Geometrical-Ring-model}
Here we extend the toy model set by studying a set of geometrical ring models  with the spin of $a= +0.5$ and various geometries for the magnetic and the velocity fields as well as the source inclination. Figure \ref{G-Ring_Model} presents $EB$-correlation phase for models with zero incliation, hereafter $\mathrm{i}$, and zero fluid speed, referred as v, (top row), models with $\mathrm{i}$=17 deg while no fluid speed (middle row) and models with $\mathrm{i}$=17 deg and non-zero fluid speed bottom row, respectively. In each row, from the left to right we study different magnetic field morphologies including a radial, toroidal and a mixed magnetic field, respectively. In each panel we overlay the ring image, on the top right corer, to build an intuition about the geometry of the ring. 

From the plot it is seen that the phase of the $EB$-correlation almost vanishes for the case with zero inclination and zero fluid speed, while it is non-zero for non-zero inclination with no fluid motion. This illustrates the importance of non-zero inclination in launching the phase. \footnote{We have also tried models with zero inclination and non-zero fluid speed and confirmed that the phase remains almost zero as well. }

Furthermore, it is inferred that different B-field geometries lead to distinct phase patterns which is also distinct for the case with zero (middle row) and nonzero (bottom row) fluid speed, respectively. Consequently, we argue that the magnetic and velocity field geometries are both important in driving the $EB$-correlation phase. In a future work we make more direct statement on the role of changing the B and v field geometries in making various non-zero phases.

\section{Numerical methodology}
\label{Numerics}
In this section, we present our numerical modeling and describe how polarized images are formed during the calculation of radiative transfer.

\begin{figure*}[th!]
\center
\includegraphics[width=\textwidth]{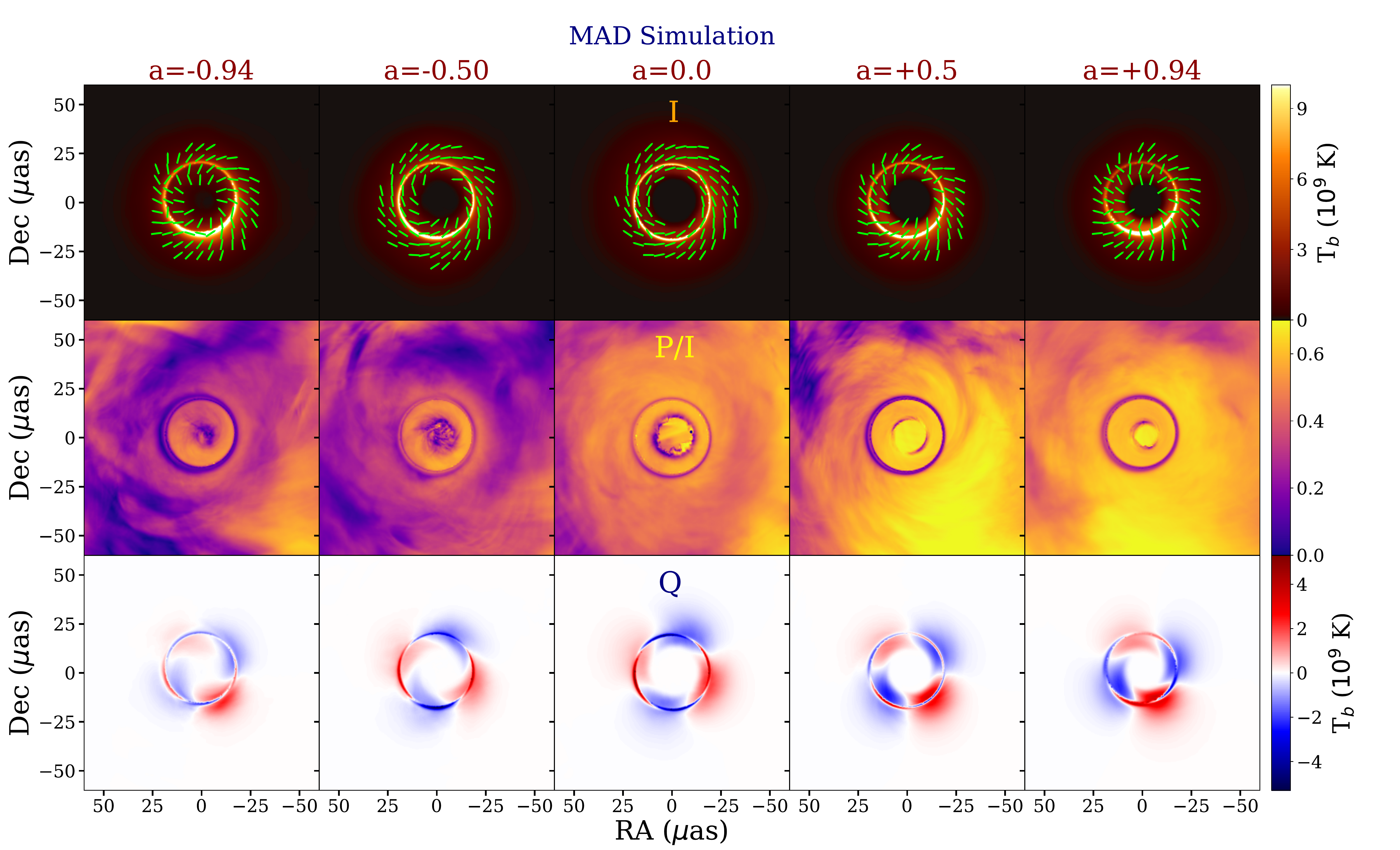}
\caption{Time averaged images of MAD simulations with different BH spins. From the top to bottom, we present the total intensity (I), fractional linear polarization (P/I), and the Stokes Q parameter. As expected images show a clear-handedness in their EVPAs.}
\label{Time-averaged-MAD}
\end{figure*}

\subsection{GRMHD simulations}
\label{GRMHD-simulations}
We employ numerical simulations in modeling the plasma flow, using an ideal and GRMHD simulation set in the Kerr metric, with BH spin taken as a free parameter \citep{1999ApJ...522..727K, GammieIHARM2003,2005ApJ...635..723A, 2007A&A...473...11D}, which is fixed at $a = (\pm 0.94, \pm 0.5, 0)$. We integrate the GRMHD equations in 3D using the H-AMR algorithm \citep{2022ApJS..263...26L}, taking into account both a magnetically arrested disk (MAD) \citep{1974Ap&SS..28...45B, 2003ApJ...592.1042I, 2003PASJ...55L..69N} as well as a standard and normal evolution (SANE) case \citep{2003ApJ...599.1238D, GammieIHARM2003, 2012MNRAS.426.3241N}, with different magnetic fluxes and different adiabatic indices: $\Gamma_{\mathrm{ad}}$ of 13/9 (MAD) and 5/3 (SANE), respectively. The initial conditions for the GRMHD simulations are taken as a torus; i.e. with a constant angular momentum \citep{1976ApJ...207..962F}, where the orbital angular momentum is parallel or antiparallel with respect to the BH spin. The initial torus is seeded by a weak and poloidal magnetic field. The particular choice of the initial torus is motivated by the emitting flow, near the event horizon, that remains in a steady state, decoupled from the flow at larger distances. Upon starting the evolution, the plasma experiences instabilities including the magnetorotational instability (MRI) \cite{1992ApJ...400..610B} as well as other instabilities such as the magnetic Rayleigh-Taylor (RT) instability \citep{2018MNRAS.478.1837M}, leading to turbulence and the emergence of a low-density, highly magnetized bubbles in MAD simulations. These play a very important role in driving angular momentum transport. These instabilities lead to an inward accretion flow of the matter directed toward the black hole. Subsequently, the plasma tends to a state with (i) a mildly magnetized midplane, (ii) a coronal component in which the gas to magnetic pressure $\beta \equiv P_g/P_B \simeq 1$, and (iii) a very strongly magnetized funnel region near the BH poles with $\sigma \equiv B^2/(4 \pi \rho c^2) \gg 1$.

\subsection{BH Imaging}
\label{imaging}
In making the BH images, we use the general relativistic radiative transfer (GRRT) algorithm, implemented in {\sc ipole} \citep{Moscibrodzka&Gammie2018}. Each image is generated using a field of view (FOV) of 200 $\mu$as with a resolution of 400 $\times$ 400 pixels. For the images, the impact of synchrotron emission, self-absorption, Faraday rotation, and Faraday conversion are all taken into account. Since the GRRT is not scale-invariant, to perform the ray-tracing, we set the characteristic length scale ($r_g$) (which is determined with the BH mass $M_{BH}$ as $L = G M_{BH} /c^2$ where $G$ and $c$ refer to the gravitational constant and the speed of light, respectively) as well as the mass-density. Throughout our analysis, we fix the source to be M87* with $M_{BH} = 6.2 \times 10^9 M_{\odot}$, located at the distance of $D = 16.9$ Mpc from an observer on earth. Furthermore, we fix the mass density by adopting the observed flux at 230 GHz to be $F_{\nu}$ = 0.5 Jy \citep{PaperIV}. Following \cite{PaperV}, we fix the source inclination at 17 (163) degrees for retrograde (prograde) spins, respectively. Finally, we rotate the simulated images to match the observed position angle of the M87* forward jet at -72 deg \citep[e.g.,][]{Walker2018}.

Since the current GRMHD simulations assume the same temperature for electrons and ions, we incorporate the electron temperature through a post-processing approach, in which the thermal equilibrium is replaced with a collisionless plasma. Consequently, the electrons and ions may have different temperatures \citep{Shapiro+1976, Narayan+1995}. Following the approach of \cite{Monika+2016, PaperV, PaperVIII}, we modulate the ratio of the ion-to-electron temperature as:
\begin{equation}
\label{electron-T}
\frac{T_i}{T_e} = R_\mathrm{high} \frac{\beta^2}{1+\beta^2} + R_\mathrm{low} \frac{1}{1+\beta^2}.
\end{equation}
where $\beta$ refers to the gas-to-magnetic pressure ratio, while  $R_\mathrm{low}$ and $R_\mathrm{high}$ are both free parameters. To reduce the scope of this analysis, we restrict these parameters to $R_\mathrm{low}=1$ and $R_\mathrm{high}=20$, respectively.

\begin{figure*}[th!]
\center
\includegraphics[width=\textwidth]{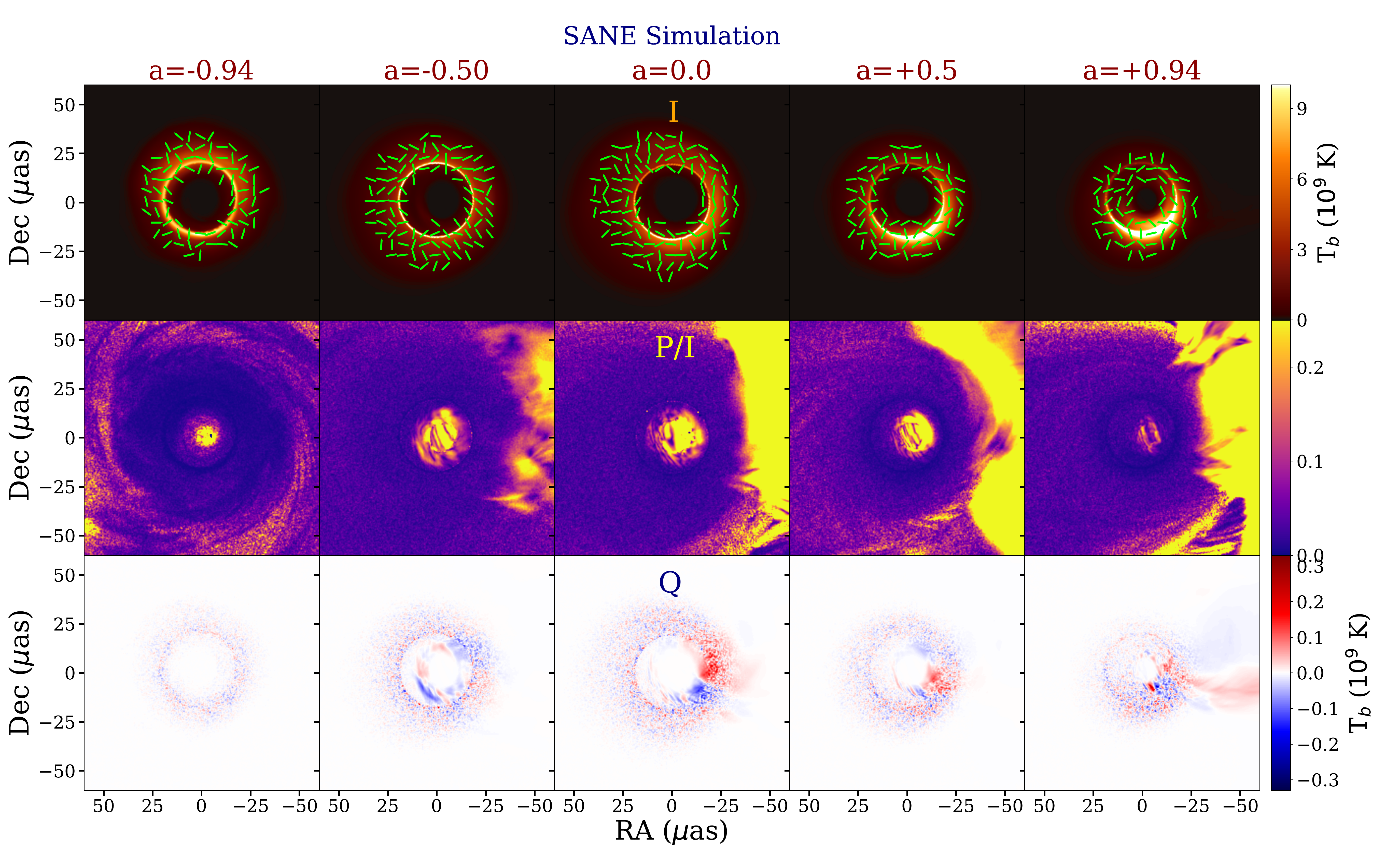}
\caption{The time-averaged images of SANE simulations with different BH spins. From top to bottom, we present the total intensity (I), fractional linear polarization (P/I), and the Stokes Q parameter. As the system is controlled by turbulence, there is not any clear-handedness in the EVPAs.}
\label{Time-averaged-SANE}
\end{figure*}

\subsection{Time-averaged BH images}
\label{time-averaged-images}
Having introduced the main set of GRMHD simulations as well as the approach for generating polarized BH images, here we present the time-averaged images of M87*. Figures \ref{Time-averaged-MAD} and \ref{Time-averaged-SANE} show the time-averaged intensity, fractional linear polarization as well as the Stokes $Q$ parameter for the MAD and SANE simulations, respectively. In each figure, from left to right, we increase the BH spin as $a = (-0.94, -0.5, 0.0, 0.5, 0.94)$. As expected, MAD simulations have more coherent EVPA patterns and clearly exhibit organized structures. Figure \ref{Time-averaged-MAD} shows that the structure of the EVPAs (at the photon ring and outside) are most similar among $a=(-0.94, +0.5, +0.94)$ versus among $a =(-0.5, 0.0)$. As we will see in the following, the phase of the $EB$-correlation function is also similar for the aforementioned cases in MAD simulations. SANE simulations, on the contrary, have mostly chaotic EVPA patterns without a clear handedness. From Figure \ref{Time-averaged-SANE}, it is inferred that their EVPAs differ substantially compared with MAD simulations. Below, we show that it also leads to substantial differences in their $EB$-correlation function. 

\subsection{Impact of Faraday Rotation}
\label{Faraday-Rotation}
\begin{figure*}[th!]
\center
\includegraphics[width=0.99\textwidth]{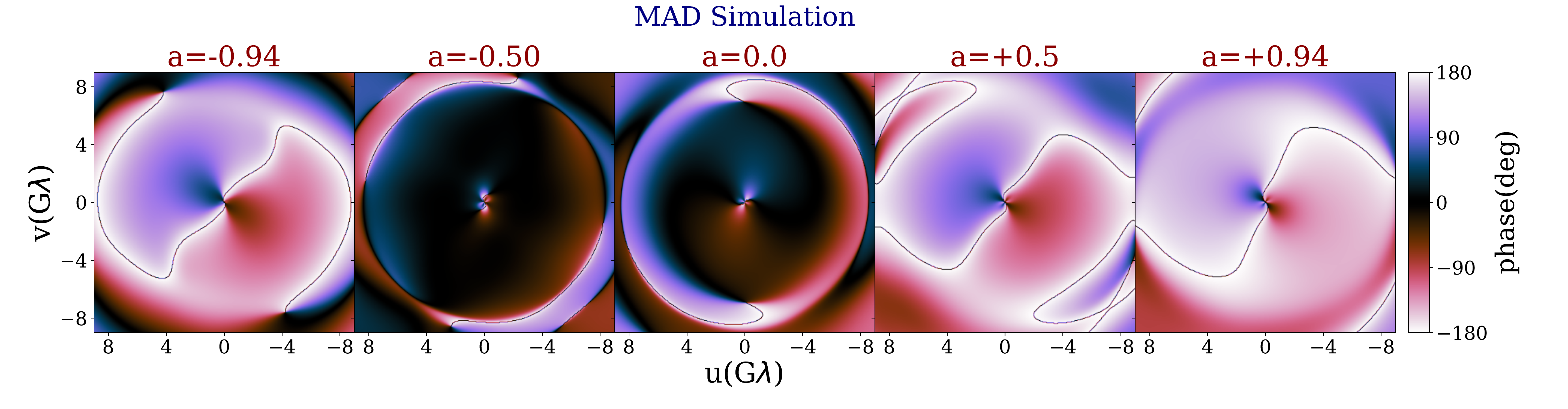}
\includegraphics[width=0.99\textwidth]{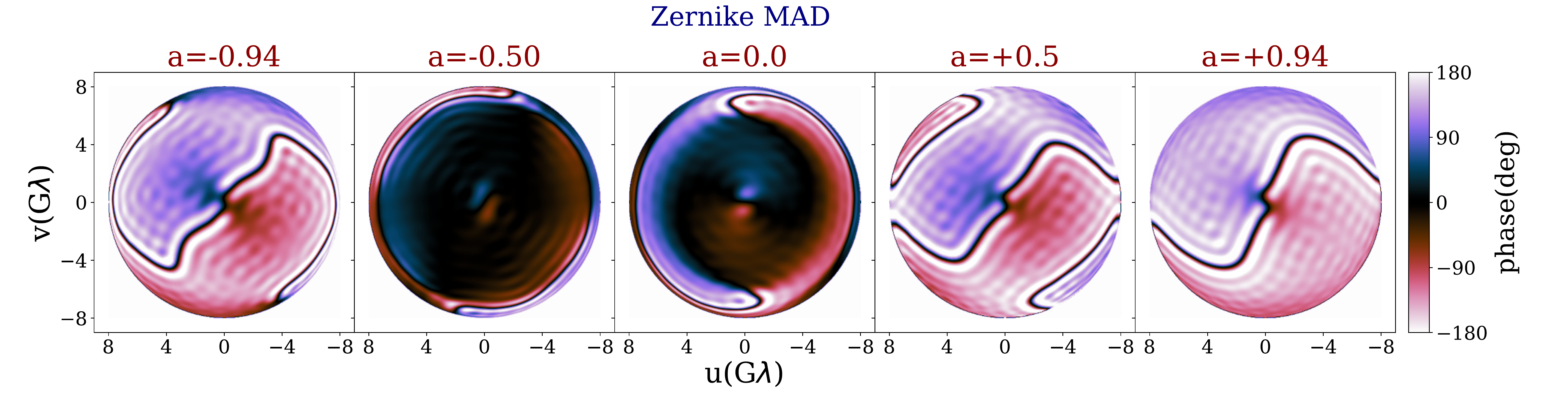}
\includegraphics[width=0.99\textwidth]{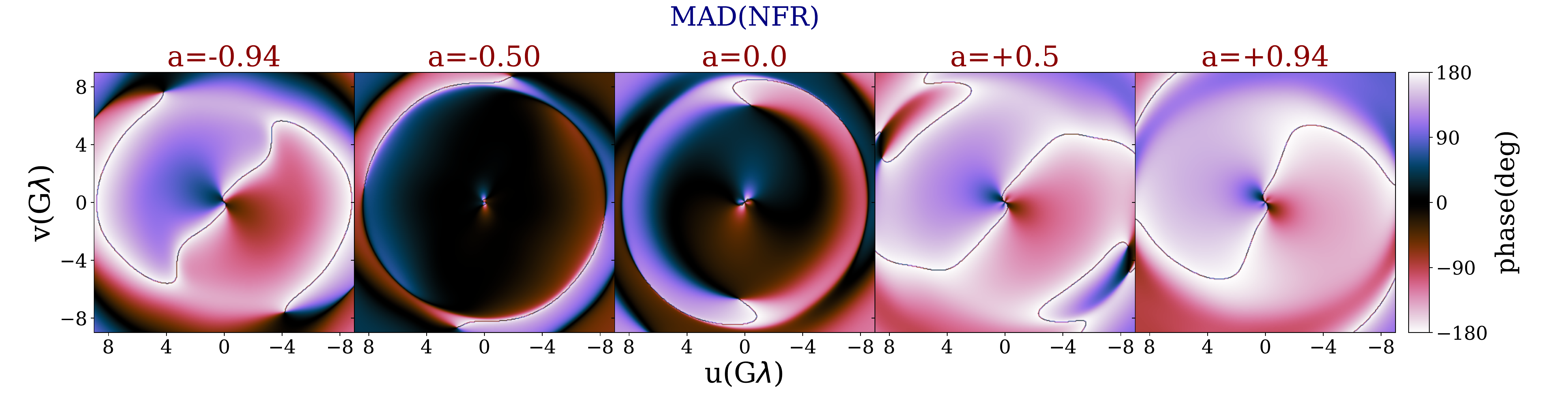}
\caption{The time-averaged phase map for the $EB$-correlation for MAD simulations (top-row), the reconstructed phase map using the Zernike polynomials (middle-row) and for MAD simulations with No Faraday Rotation (NFR) (bottom-row). To infer the phase map we adopt the $M_1$ method presented in Section \ref{phase-1st-method} in which we first compute the time-averaged $E$ and $B$ modes and then compute the correlation phase using these averaged quantities. The Zernike expansion is truncated at $n=40$ order. As the Zernike polynomials are defined on a unit circle, we have multiplied them to 8 to be similar to the original phase map from the GRMHD simulations.}
\label{MAD-average-map-1st}
\end{figure*}

\begin{figure*}[th!]
\center
\includegraphics[width=0.99\textwidth]{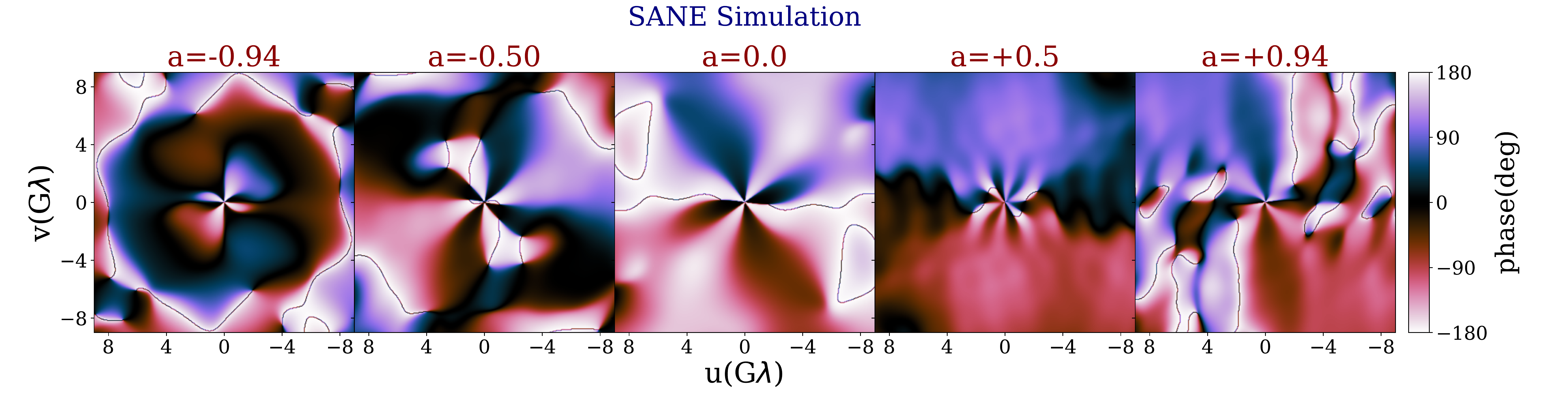}
\includegraphics[width=0.99\textwidth]{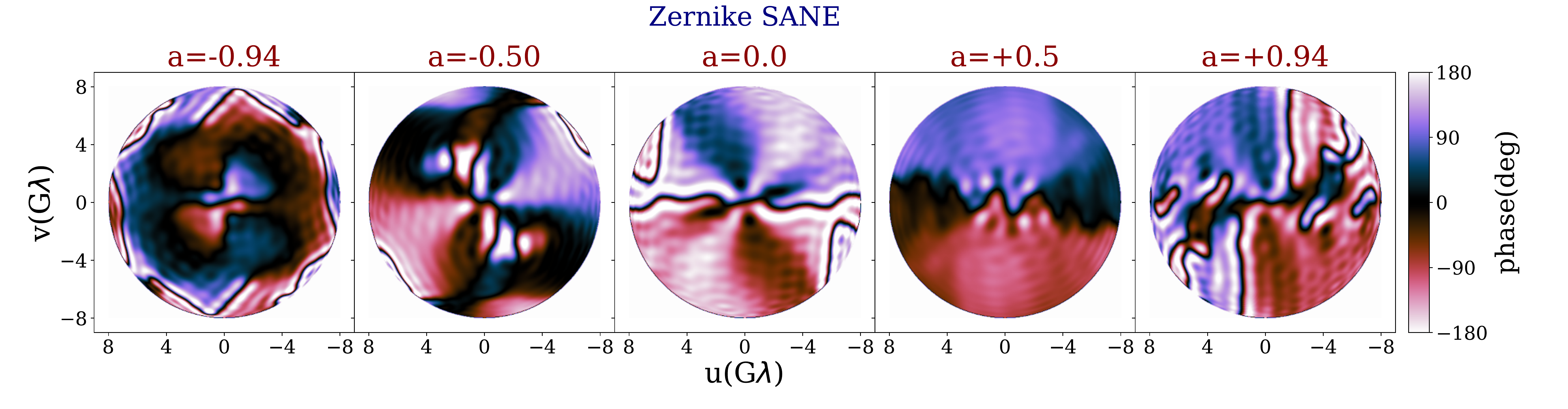}
\includegraphics[width=0.99\textwidth]{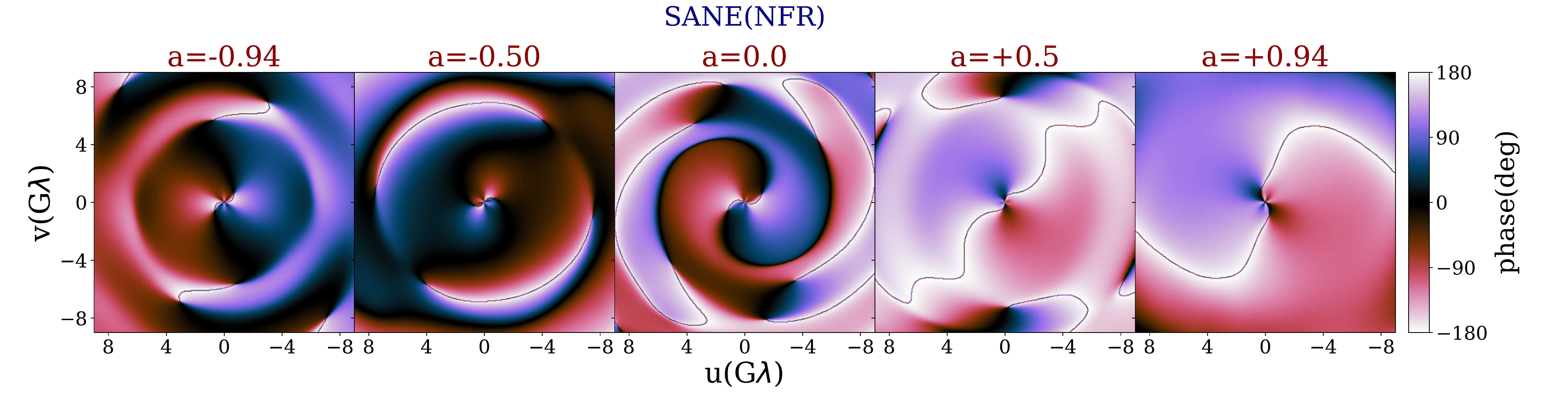}
\caption{The time-averaged phase map for the $EB$-correlation for SANE simulations (top-row), the reconstructed phase map using the Zernike polynomials (middle-row) and for SANE simulations with No Faraday Rotation (NFR) (bottom-row). To infer the phase map we adopt the $M_1$ method presented in Section \ref{phase-1st-method} in which we first compute the time-averaged $E$ and $B$ modes and then compute the correlation phase using these averaged quantities. The Zernike expansion is truncated at $n=40$ order. As the Zernike polynomials are defined on a unit circle, we have multiplied them to 8 to be similar to the original phase map from the GRMHD simulations.}
\label{SANE-average-map-1st}
\end{figure*}

In the presence of a magnetized plasma, linear polarization is altered owing to the Faraday rotation with an amount depending on the intervening plasma density, path length, magnetic field, and the temperature. To assess the impact of the Faraday rotation on the phase of the $EB$-correlation function, we infer images with the Faraday rotation switched off (hereafter NFR), where we turn off the Faraday Rotation coefficient, $\rho_V=0$, and compute the time-averaged phase of the $EB$-correlation while comparing different models. The third row in Figures \ref{MAD-average-map-1st} and \ref{SANE-average-map-1st} present the time-averaged phase map for the NFR case in MAD and SANE simulations, respectively. It is inferred from the figures that while the Faraday Rotation does not have any significant contributions in MAD simulations, it plays an important role for SANEs. 

\begin{figure*}[th!]
\center
\includegraphics[width=0.99\textwidth]{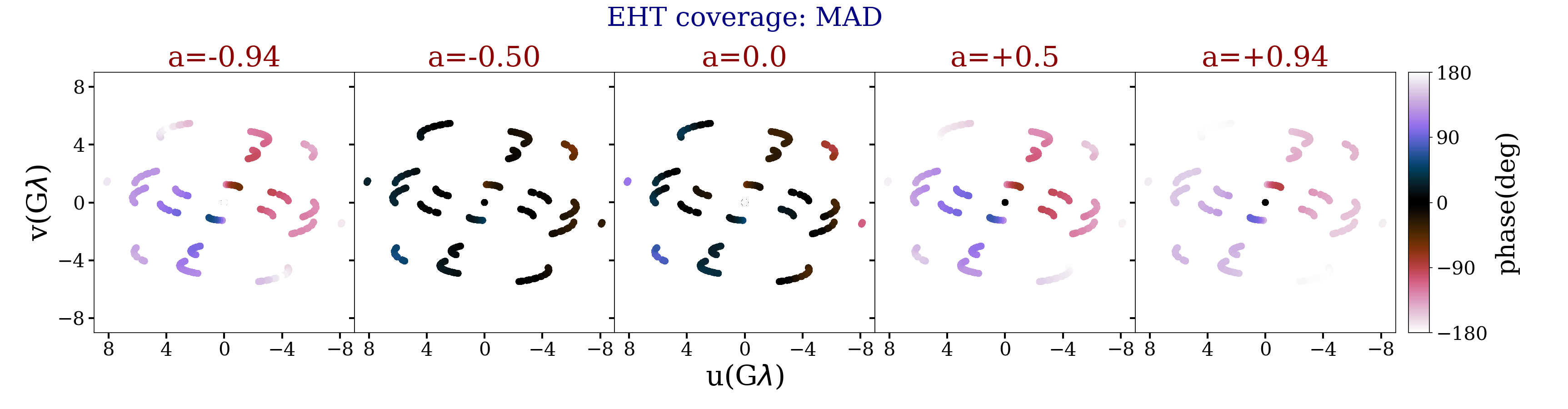}
\includegraphics[width=0.99\textwidth]{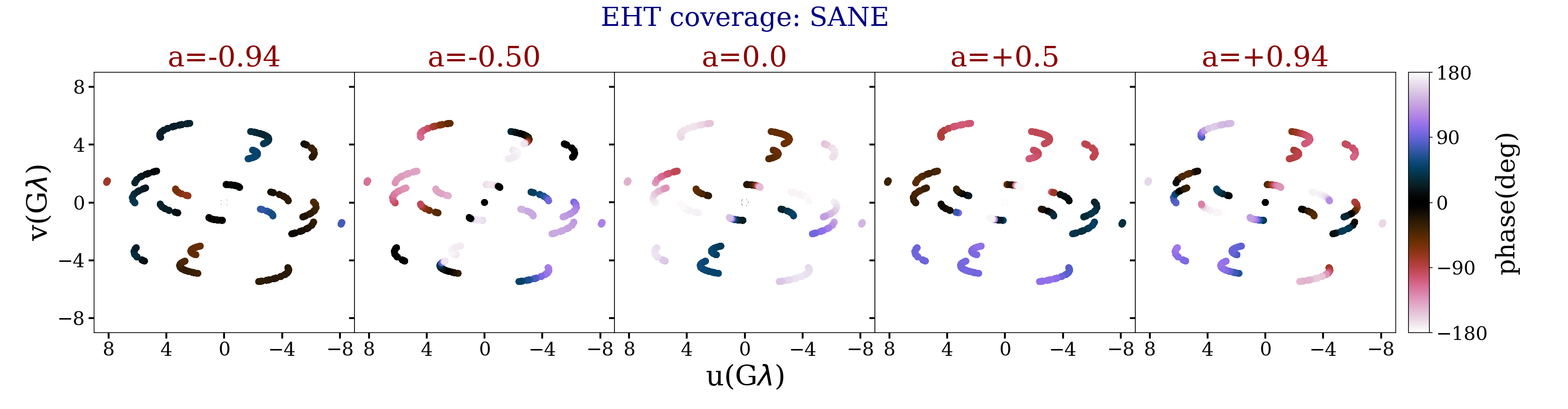}
\caption{The phase map for the $EB$-correlation function in the visibility space from the time-averaged GRMHD simulations using the EHT 2017 coverage for M87*. In the top(bottom) panel, we have used the MAD(SANE) simulations to make the synthetic data.}
\vspace{10pt}
\label{Phase-EHT-Coverage}
\end{figure*}

\section{Zernike polynomials}
\label{ZP-expansion}
Here we present a new algorithm to reconstruct $EB$-correlation maps in visibility space and show how to extract the BH features from our proposed $EB$-correlation phase map. We employ the Zernike polynomials \citep{1976JOSA...66..207N} in a unit circle as a complete and orthogonal basis for the expansion. This method is being widely used in adaptive optics \citep{2011JMOp...58.1678L} to extract features from an image. Recently, it has also been used in the galaxy cluster community \citep{2021MNRAS.503.6155C}. In our analysis, we follow the convention in \cite{1976JOSA...66..207N} for the positive and negative Zernike polynomials as: 
\begin{equation}
\label{Zernike}
\begin{split}
Z_n^m(\rho, \theta) ~=& ~
 N_n^m R_n^m(\rho) \cos{m \theta}, \\
Z_n^{-m}(\rho, \theta) ~ =& ~
 N_n^m R_n^m(\rho) \sin{m \theta} .
\end{split}
\end{equation}
In Eq. (\ref{Zernike}) only the absolute value of $m$ appears. More explicitly, we have $N_n^m = N_n^{|m|}$ and $R_n^m = R_n^{|m|}$ and the same in the $\sin{m \theta} = \sin{|m| \theta}$. Furthermore, the normalization factor is given by:
\begin{equation}
\label{norm}
N_n^m  = \sqrt{\frac{2(n+1)}{1 + \delta_{m0}}} .
\end{equation}
Also, the Zernike radial term is expressed as:
\begin{equation}
\label{radial-zernike}
R_n^m(\rho) = \sum_{s=0}^{(n-m)/2} \frac{(-1)^s (n-s)!}{s! \left( \frac{n+m}{2} - s \right)! \left( \frac{n-m}{2} - s \right)!} \rho^{n-2s},
\end{equation}
where $ 0 \leq \rho \leq 1$ is the normalized radial distance, and the azimuthal angle lies in the range $0 \leq \theta \leq 2\pi$. Zernike polynomials are only non-zero for $n-m = \mathrm{even}$. We may express any arbitrary function $\phi(\rho, \theta)$ as the weighted sum of Zernike polynomials as: 
\begin{equation}
\label{phi-expansion}
\phi(\rho, \theta) = \sum_{n=0}^{\infty} \sum_{m=0}^{n} c_{nm} Z_{n}^{m}(\rho, \theta),
\end{equation}
where $c_{nm}$ refers to the Zernike moments, given by:
\begin{equation}
\label{Zernike-Moment}
c_{nm} = \left(\frac{1}{\pi} \right) \int_{0}^{1} \int_{0}^{2 \pi} \phi(\rho, \theta) Z_{n}^{m}(\rho, \theta) \rho d \rho d \theta.
\end{equation}
In the above expansion, we have used the orthogonality of the Zernike polynomials:
\begin{equation}
\label{Orthogonal}
\int_{0}^{1} \int_{0}^{2\pi} Z^{m}_{n}(\rho, \theta) Z^{m'}_{n'}(\rho, \theta) \rho d \rho d \theta = \pi \delta_{n n'} \delta_{m m'} .
\end{equation}

Below, we use the Zernike polynomials as a novel way to reconstruct and extract the features from the $EB$-correlation phase map. We explicitly show that the Zernike polynomials might be very useful in distinguishing features from maps that are qualitatively similar to one another. The generality of these patterns must be checked against a wider library of images and is left to a future study. 

\section{Phase correlation Analysis}
\label{Correlation-Phase}
In this section, we compute the $EB$-correlation phase using the synthetic data generated from the time-averaged GRMHD simulations and by adopting the EHT 2017 coverage  for M87* \citep{PaperVII}. In our analysis, we investigate three different cases. First, we calculate the $EB$-correlation phase using the discrete EHT coverage. We then extend this analysis to grids in the $uv$-space. Finally, while in the first two cases, we use the $M_1$ method in Section \ref{phase-1st-method}, as a consistency check, we also perform the analysis using the $M_2$ method in Section \ref{phase-2nd-method}. 

\subsection{$EB$-correlation phase: Discrete EHT coverage}
\label{EHT-discrete-uvfits}
We infer the $EB$-correlation phase map for the time-averaged GRMHD simulations using the EHT 2017 coverage \citep{PaperVII}. In our analysis, we follow the $M_1$ method described in Section \ref{phase-1st-method}. 

Figure \ref{Phase-EHT-Coverage} presents the $EB$-correlation phase map for the synthetic observational data made using the time-averaged images of GRMHD simulations, adopting the EHT 2017 coverage for the M87* source. Despite the sparse $uv$ coverage, distinct patterns are visible from different columns. This motivates us to perform an in-depth analysis making some grids in the visibility space. 

\subsection{$EB$-correlation phase: EHT gridded coverage}
\label{grid-based}
To get a better sense of different patterns in the $EB$-correlation phase map as well as to circumvent the sparse $uv$-coverage in the EHT data, here we use a grid of $uv$-space coverage to infer the time-averaged $E$ and $B$ modes. We continue using the $M_1$ method in Section \ref{phase-1st-method} to calculate the time-averaged correlation phase. 

\begin{figure*}[th!]
\center
\includegraphics[width=0.98\textwidth]{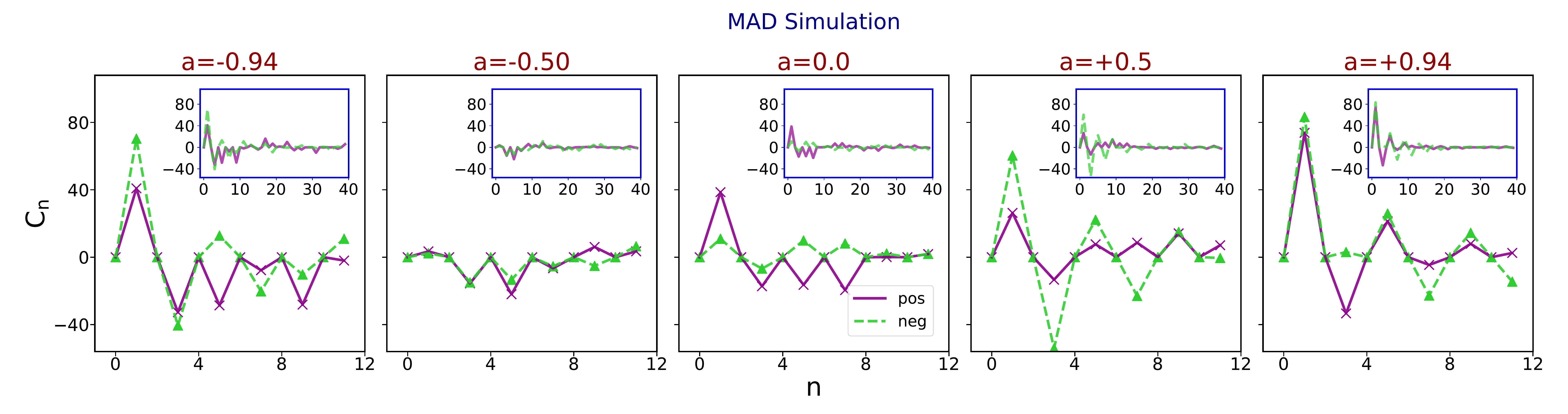}
\includegraphics[width=0.98\textwidth]{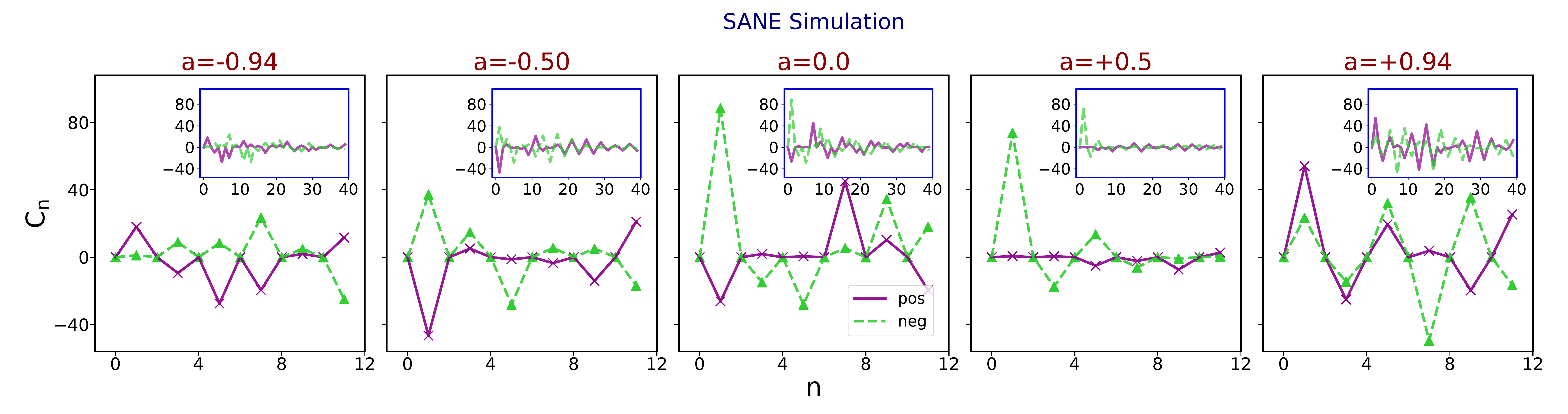}
\caption{The Zernike moments vs. the $n$ index for MAD (top) and SANE simulations (bottom). From left to the right, we increase the BH spin $a = (-0.94, -0.5, 0.0, +0.5, +0.94)$. In each panel, we sum over all of the non-zero $m$ indices for a given $n$ splitting the positive and negative terms with the solid-magenta and dashed-green lines, respectively. Different simulations show distinct patterns in $C_n$ (as defined in Eq. \ref{Cn}) vs. $n$ that can eventually be used to distinguish them from one another. In each panel, the subplot refers to the case with an extended value of $n$s, indicating that in most cases we do not need to consider n-orders above 20.}
\label{Zernike_Moments_norder}
\end{figure*}

\begin{figure*}[th!]
\center
\includegraphics[width=0.98\textwidth]{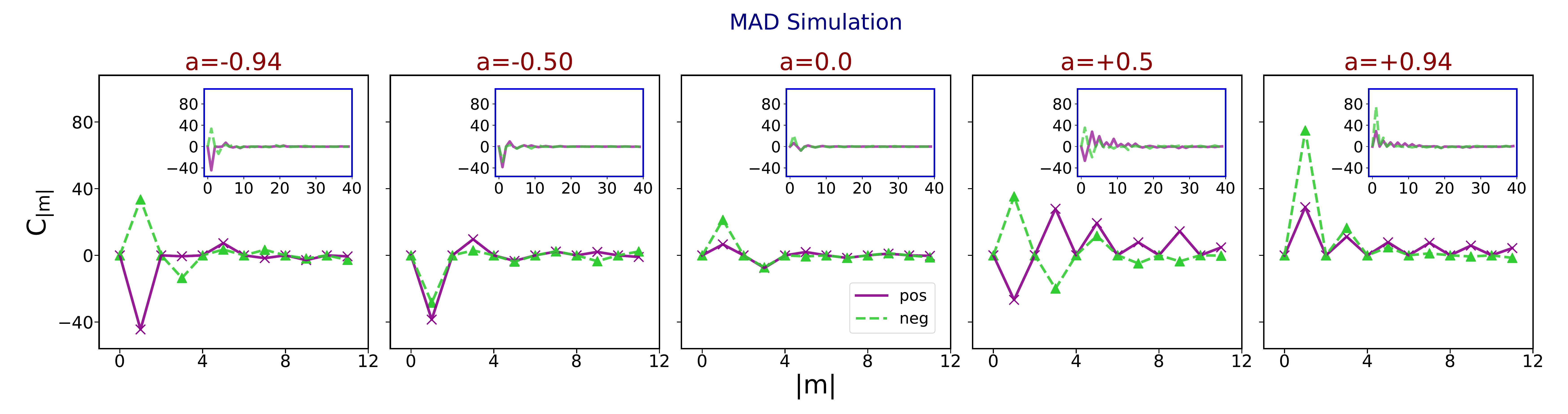}
\includegraphics[width=0.98\textwidth]{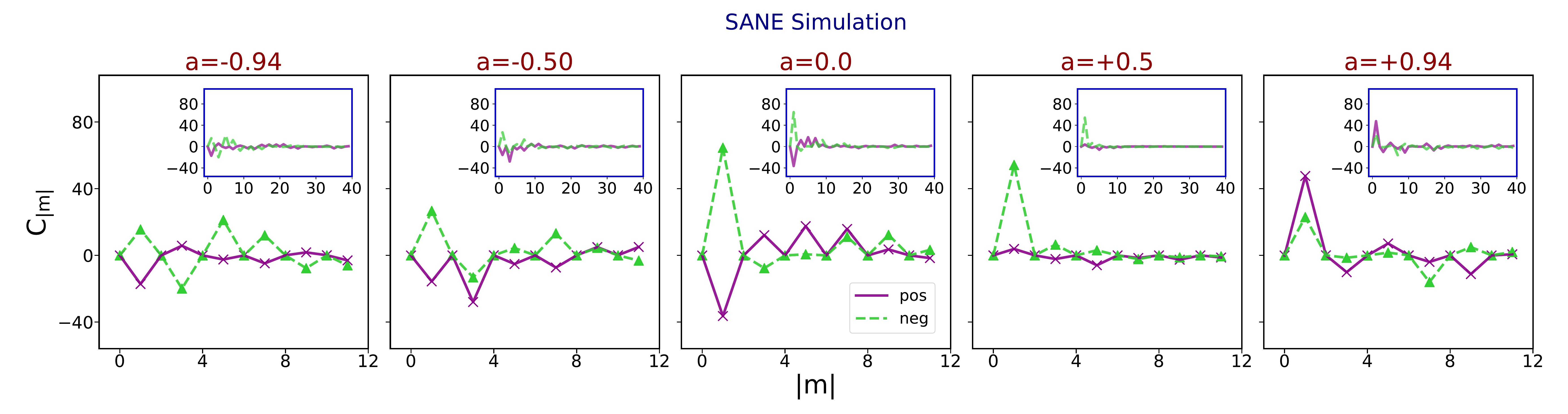}
\caption{The Zernike moments vs. the absolute value of the $m$ index for MAD (top) and SANE (bottom) simulations. From left to the right, we increase the BH spin in $a = (-0.94, -0.5, 0.0 +0.5, +0.94)$. In each panel, we compute the summed $C_{nm}$ for all $n$ indices that include the associated $m$ index, splitting the positive and the negative terms with solid-magenta and dashed-green lines, respectively. Different simulations show different patterns in their $C_{|m|}$ (as defined in Eq. \ref{Cm}) vs $|m|$ that might be used to distinguish them from one another. In each panel, the subplot refers to the case with an extended value of $|m|$s that again indicate we do not need to consider $m$ orders above 20. }
\label{Zernike_Moments_morder}
\end{figure*}

Figures \ref{MAD-average-map-1st} and \ref{SANE-average-map-1st} present the $EB$-correlation phase map for time-averaged simulations with different BH spins. In each figure, from the top to bottom, different rows present the original simulation (MAD and SANE), the reconstructed phase maps using the Zernike polynomials, and the phase map for the case with No Faraday Rotation (NFR). In each row, from the left to right, we increase the BH spin in $a = (-0.94, -0.5, 0.0, +0.5, +0.94)$. 

A dipolar pattern is visible in all different maps owing to the conjugate symmetry in a baseline between the two stations
$i$ and $j$. More explicitly, it can be shown that $E^*_{ij} = E_{ji}$ and $B^*_{ij} = E_{ji}$, see Eq. 7 of \cite{2021ApJ...910L..12E} for more details. Consequently, the phase flips under $ [u,v] \rightarrow [-u, -v]$. However, it does not lead to any flips when we only flip either of $u$ or $v$ values in the visibility space. 

Focusing on MAD simulations, there are two distinct types of phase maps. In the first case, which includes $a = (-0.94, +0.5, +0.94)$, the phase map contains visibly distinct plus and minus sign phases, which are mainly centered around $\theta \geq 90$ deg. Comparing the phase map with the EVPA patterns from Figure \ref{Time-averaged-MAD}, it is observed that there are some similarities between their EVPAs as well. In the second case, which includes $a = (-0.5, 0.0)$, the phase map is mainly centered on rather small values, $\theta \leq 30$ deg. Comparing this with the EVPA patterns from Figure \ref{Time-averaged-MAD}, it is observed that their EVPAs show very similar patterns as well. 

For the SANE simulations, in the third-fourth rows, it is clearly seen that various spins behave rather differently. Comparing this with the EVPA patterns in Figure \ref{Time-averaged-SANE}, it is observed that they have very different patterns as well. We speculate that it might owe to the turbulence in the SANE simulations. 

\begin{figure*}[th!]
\center
\includegraphics[width=0.99\textwidth]{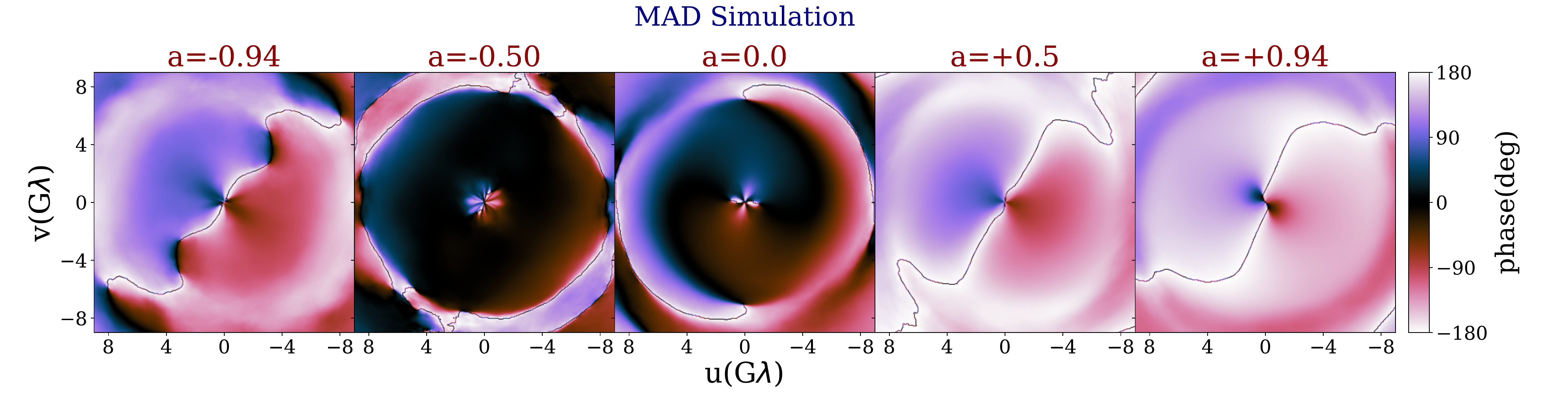}
\includegraphics[width=0.99\textwidth]{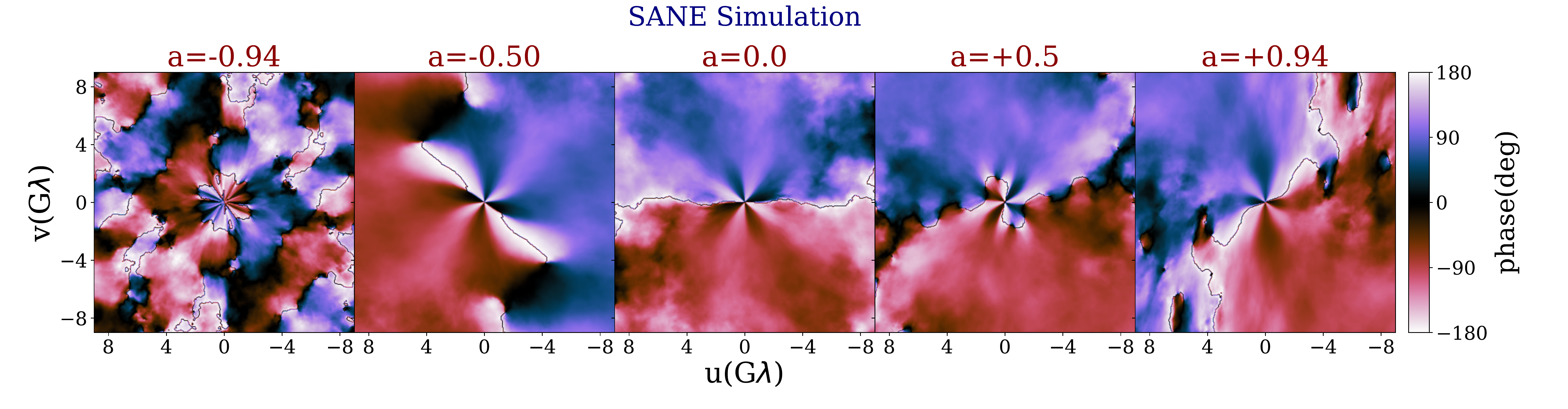}
\caption{The time-averaged phase map for the $EB$-correlation for MAD and SANE simulations using the $M_2$ method described in Section \ref{phase-2nd-method}, where we first compute the $EB$-correlation at every snapshot and then calculate the time-averaged phase map. Comparing the results with Figures  \ref{MAD-average-map-1st} and\ref{SANE-average-map-1st},
it is seen that both maps are qualitatively similar.}
\label{MAD-SANE-rho-average-map}
\end{figure*}

\begin{figure*}[th!]
\center
\includegraphics[width=0.99\textwidth]{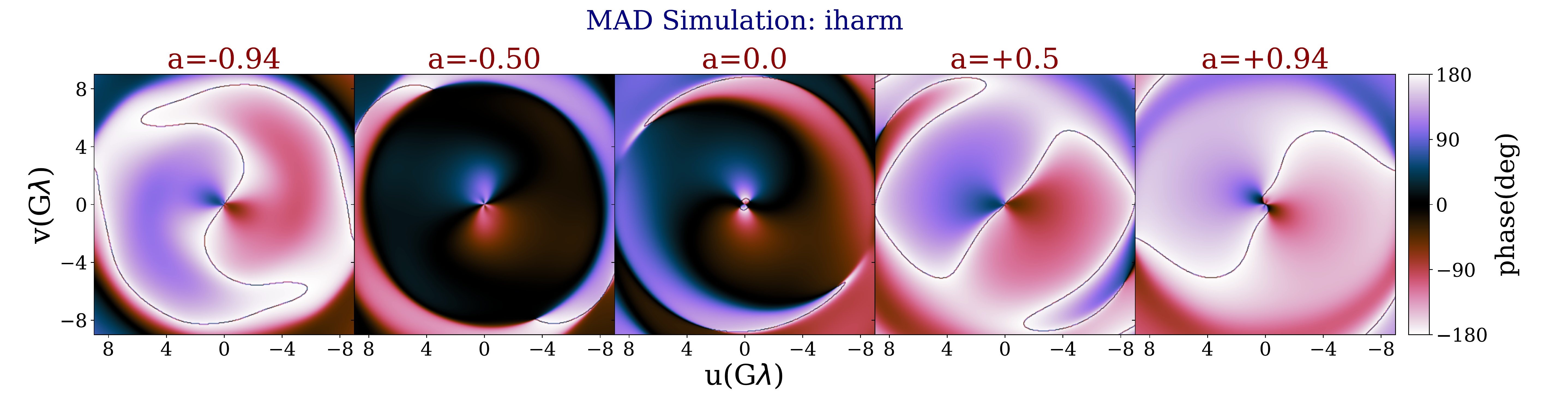}
\caption{The time averaged phase map for the $EB$-correlation in MAD simulations from the GRMHD simulations of {\sc iharm}. To infer the phase map we follow $M_1$ method described in Section \ref{phase-1st-method}. It is clearly seen that the phase map is very similar to that of the MAD simulations from H-AMR simulations, which demonstrates that the observed patterns in MAD simulations are robust and rather independent of initial conditions. }
\label{MAD-Simulation-iharm}
\end{figure*}

Comparing the original phase maps, from the first row, with the reconstructed ones from the Zernike polynomials, as shown in the second row, in Figures \ref{MAD-average-map-1st} and \ref{SANE-average-map-1st}, it is demonstrated that the Zernike polynomials do a good job at reconstructing the main features in the phase map. Consequently, in the following, we will use this basis to make more quantitative comparisons between the phase maps from different simulations. 

\subsubsection{Feature Extraction from the phase map using the Zernike polynomials}
\label{Feature-Extraction}
Here we use the reconstructed $EB$-correlation phase map from Zernike polynomials and extract BH features using two distinct empirical metrics. 

$\blacksquare$ As the first metric, for every index $n$, we sum over the $m$ indices that appeared in the Zernike Moment $c_{nm}$, splitting the positive (pos) and the negative (neg) $m$ indices:
\begin{equation}
\label{Cn}
C_n  \equiv \sum_{m} c_{nm}  ~~~~~, ~~~~~ m = [\mathrm{pos}, \mathrm{neg}].
\end{equation}
Figure \ref{Zernike_Moments_norder} presents $C_n$ for MAD and SANE simulations. In each panel, we present the $C_n$ vs. $n$ index in the range $0 \leq n \leq 12$, as key players in the expansion. To illustrate the impact of higher index $n$s, in each panel, we add a sub-panel with $n \leq 40$. Finally, we illustrate the positive and negative $m$ indices with solid-magenta and dashed-green lines, respectively. From the plot, it is evident that simulations with different BH spin as well as accretion types are distinct based on their $C_n$ profiles, which also differ between the positive and negative $m$ indices. Furthermore, in most cases, higher $n$ orders are suppressed compared with the lower values (with an exception of a SANE simulation with $a = + 0.94$). More interestingly, it is seen that using the Zernike moments we may break some of the degeneracies between the cases that are qualitatively very similar. For instance, the positive $m$ index for $a = -0.5$ in MAD simulations is not the same as for MAD with $a = 0.0$. The same is true between MADs with spin $a = -0.94, +0.5, +0.94$, which were qualitatively very similar. Consequently, we argue that the $C_n$ vs. $n$ diagram might be very useful in distinguishing different cases. The validity of this statement must be checked against different electron temperature profiles and is left to a future study. 

\begin{figure*}[th!]
\center
\includegraphics[width=0.99\textwidth]{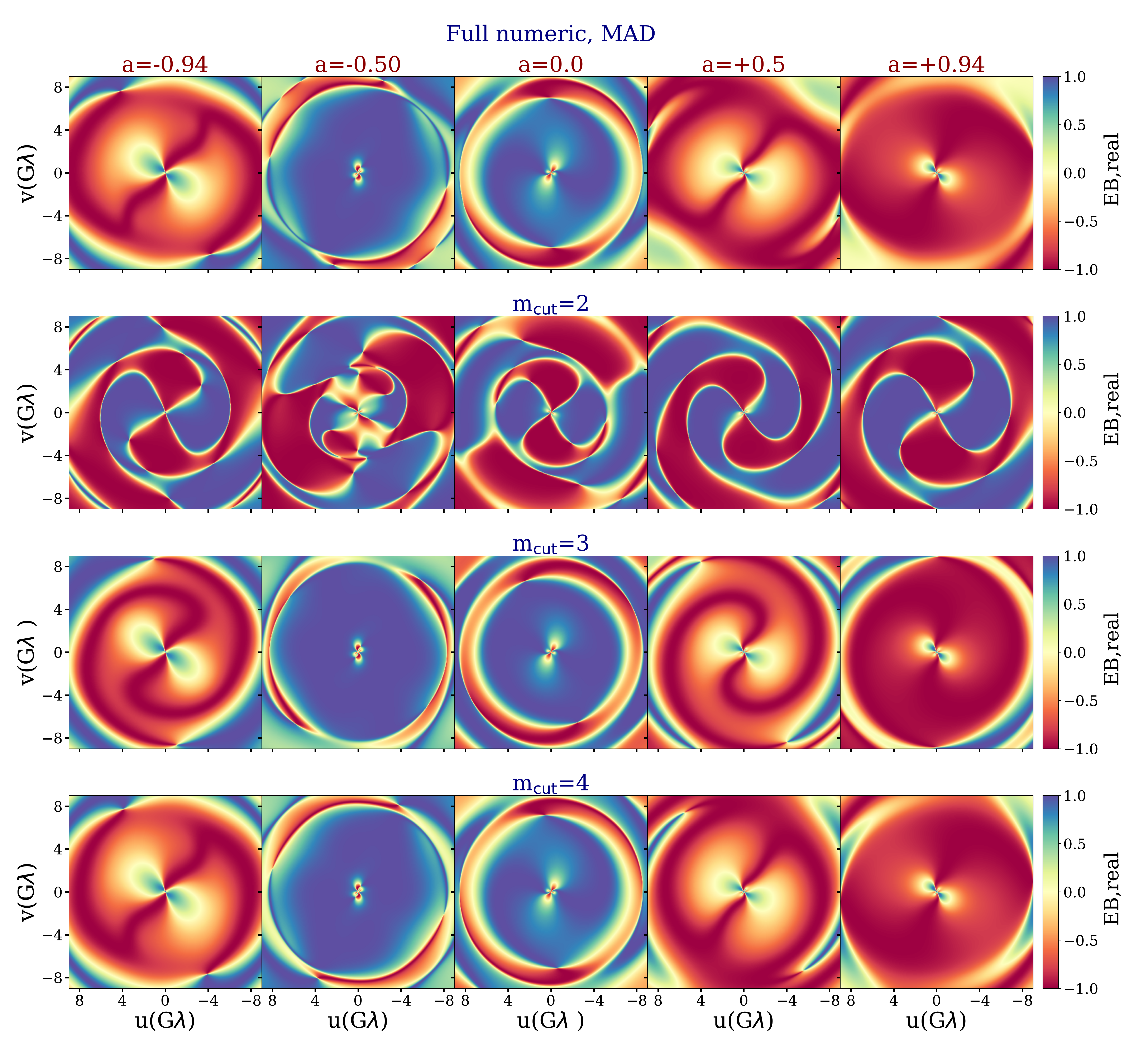}
\caption{The comparison between the real part of the $EB$-correlation in MAD simulations from the full-numeric as well as the $m$-order cut. From the top to bottom, we present the full numeric, the case with  $m_{\mathrm{cut}}$ =2,  $m_{\mathrm{cut}}$ =3 and  $m_{\mathrm{cut}}$ =4, respectively.  }
\label{Comparison_EB-Real_MAD}
\end{figure*}
$\blacksquare$ As the second metric, we sum over all $n$ indices that include a given $m$ index, splitting out the positive (pos) and negative (neg) terms:
\begin{equation}
\label{Cm}
C_{|m|}  \equiv \sum_{n} c_{nm}  ~~~~~, ~~~~~ m = [\mathrm{pos}, \mathrm{neg}].
\end{equation}
Figure \ref{Zernike_Moments_morder} presents $C_{|m|}$ vs. $|m|$ index from different simulations. In all cases, the $C_{|m|}$ is suppressed at higher $|m|$ values.
Furthermore, each simulation has its own pattern which differs from other cases. This is also true for degenerate cases. Combining this with the results from Figure \ref{Zernike_Moments_norder}, it is observed that degeneracies between various models are broken if we use the Zernike polynomials. This suggests that the combination of these two metrics might be very informative in quantifying structural differences between various cases.  

\subsection{$EB$-correlation: Second approach}
\label{second-approach-phase}
So far we have only computed the $EB$-correlation phase map using the $M_1$ method from Section \ref{phase-1st-method}. Here we make a consistency check by computing the phase map using the $M_2$ method as described in Section \ref{phase-2nd-method}. Figure \ref{MAD-SANE-rho-average-map} presents the map using the $M_2$ method. Comparing the map with the one from Figures \ref{MAD-average-map-1st} and \ref{SANE-average-map-1st}, it is evident that they are quite similar. Motivated by this, in the remainder of the paper we focus on the $M_2$ method. 

\begin{figure*}[th!]
\center
\includegraphics[width=0.99\textwidth]{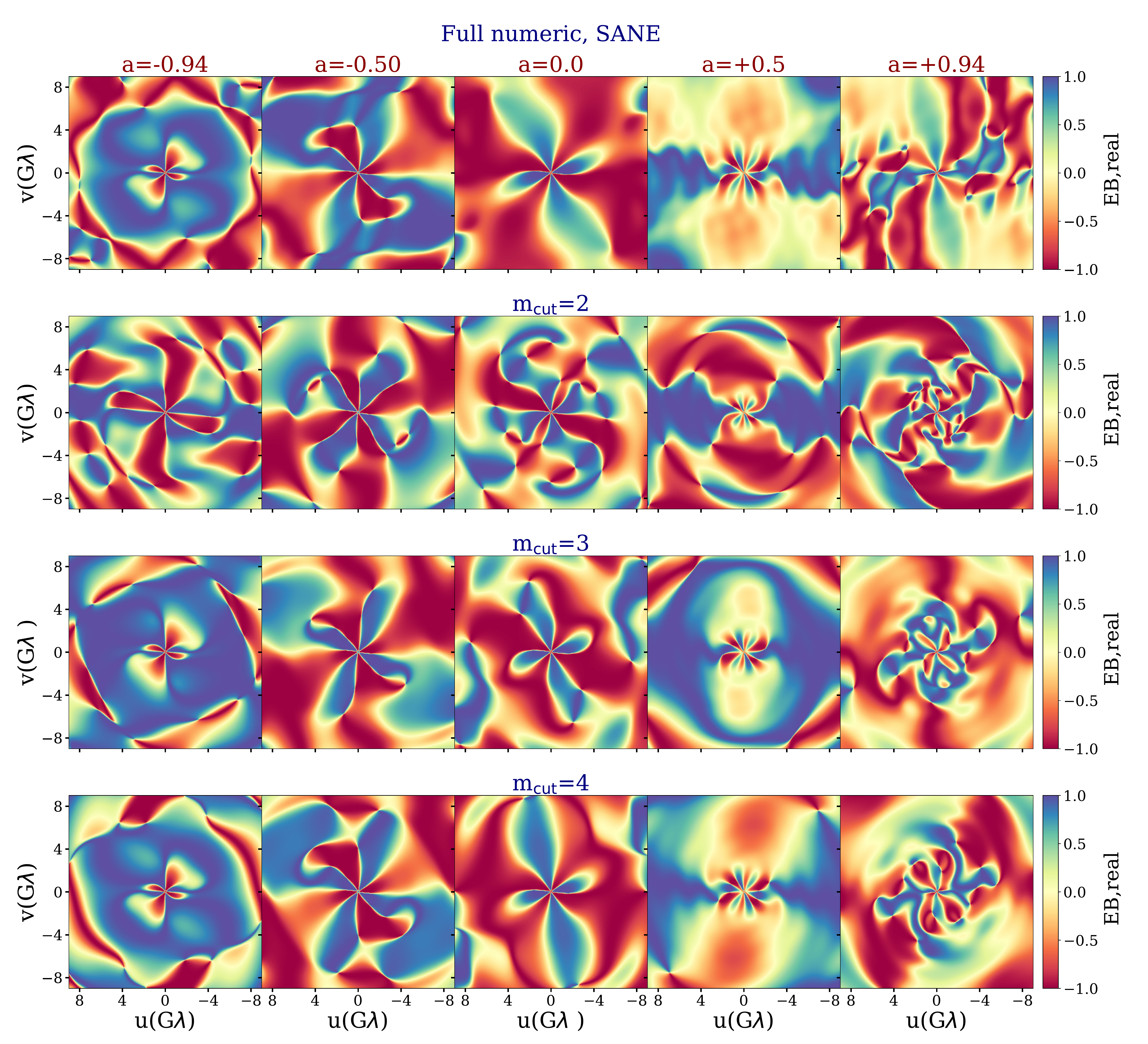}
\caption{Comparison between the real part of the $EB$-correlation in SANE simulations from the full-numeric as well as the $m$-order cut. From the top to bottom, we present the full numeric, the case with  $m_{\mathrm{cut}}$ =2,  $m_{\mathrm{cut}}$ =3 and  $m_{\mathrm{cut}}$ =4, respectively. }
\label{Comparison_EB-Real_SANE}
\end{figure*}
\subsection{Comparison with {\sc iharm} simulation}
\label{Comparison}
Here we make a comparison between the $EB$-correlation phase map inferred using the H-AMR simulation, as done in this work, and the one from the standard library of GRMHD simulations of {\sc iharm} by \cite{GammieIHARM2003,2021JOSS....6.3336P, Emami2022}. Figure \ref{MAD-Simulation-iharm} presents the phase map for the time-averaged GRMHD simulations of {\sc iharm}. The phase map in the MAD case is very similar to the one from H-AMR. However, it differs for SANE simulations as the level of the turbulence is different between the {\sc iharm} and H-AMR simulations. Consequently, to avoid any confusions, we skip making the comparison for SANE simulations. 

\section{Semi-analytic estimation of the $EB$-correlation function}
\label{morder-expansion}
Having presented the general approach in computing the $\rho_{\mathcal{EB}}(u,v)$, here we follow a semi-analytical algorithm from \cite{PaperVIII} and make an $m$-order expansion for the $E$ and $B$ modes, in the visibility space. We then compare that with the full numerical results as inferred from Section \ref{grid-based}.

In this method, we expand the $E(u,v)$ and $B(u,v)$ modes as: 
\begin{equation}
\label{E-B-Fourier}
\begin{split}
E(u,v) =& \sum_
{m=-\infty}^{\infty} i^{-m} Re \bigg{[} e^{i(m-2) \arctan{(u/v)}} W_m(\sqrt{(u^2 + v^2)}) \bigg{]}, \\
B(u,v) =& \sum_
{m=-\infty}^{\infty} i^{-m} Im \bigg{[} e^{i(m-2) \arctan{(u/v)}} W_m(\sqrt{(u^2 + v^2)}) \bigg{]},
\end{split}
\end{equation}
where $W_m(X)$ is defined as:
\begin{align}
\label{W-m-u}
W_m(X) \equiv & \int_{\rho_{\mathrm{min}}}^{\rho_{\mathrm{max}}} \int_{0}^{2 \pi} \bigg{(} Q(\rho, \phi) + i U(\rho, \phi) \bigg{)} J_{m}(2 \pi \rho X) \nonumber \\
& ~~~ \times e^{-i m \phi} \rho d \rho d \phi.
\end{align}
Using Eq. (\ref{E-B-Fourier}) and following the $M_1$ method in Section \ref{phase-1st-method}, we first compute the time-averaged $E$ and $B$ modes and then compute the $EB$-correlation function. However, since the $E$ and $B$ expansions in Eq. (\ref{E-B-Fourier}) includes infinite terms, we must truncate the expansion up to a specific order. Former literature \cite[see for example][]{PaperVIII} has only considered the linear polarization at $m_{\mathrm{cut}}=2$ order. Here we go several steps beyond this approximation, incorporating the new terms up to $m_{\mathrm{cut}}=\pm 4$ order, and explicitly check out whether the above expansion is sufficient at recovering pattern morphology. As we will show, the $EB$-correlation function is sensitive to these higher-order terms in the expansion. In Eq. (\ref{E-B-expansion}), we expand the $E$ and $B$ modes and compute the real and imaginary parts of the $EB$-correlation function in Eqs. (\ref{E-B-rr}) and (\ref{E-B-im}), respectively. It is evident that while the real part of the $EB$-correlation is sensitive to even-even and odd-odd terms, the imaginary part of the correlation function is only sensitive to the even-odd terms and so the former approach adopted in \citet{PaperVIII} does not work here. 

\begin{figure*}[th!]
\center
\includegraphics[width=0.99\textwidth]{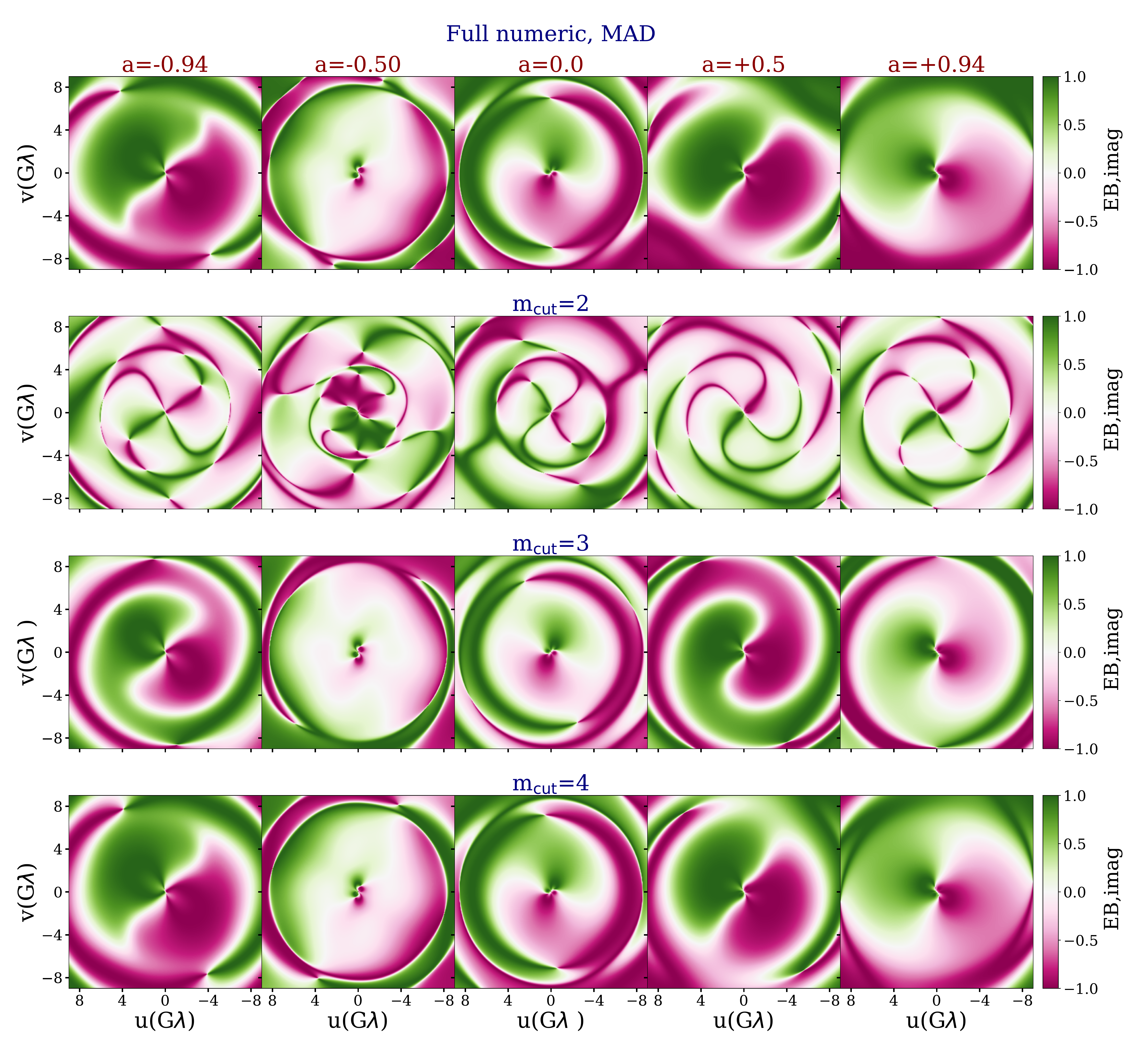}
\caption{The comparison between the imaginary part of the $EB$-correlation in MAD simulations from the full-numeric as well as the $m$-order cut. From the top to bottom, we present the full numeric, the case with  $m_{\mathrm{cut}}$ =2,  $m_{\mathrm{cut}}$ =3 and  $m_{\mathrm{cut}}$ =4, respectively. }
\label{Comparison_EB-Imagin_MAD}
\end{figure*}

\begin{figure*}[th!]
\center
\includegraphics[width=0.99\textwidth]{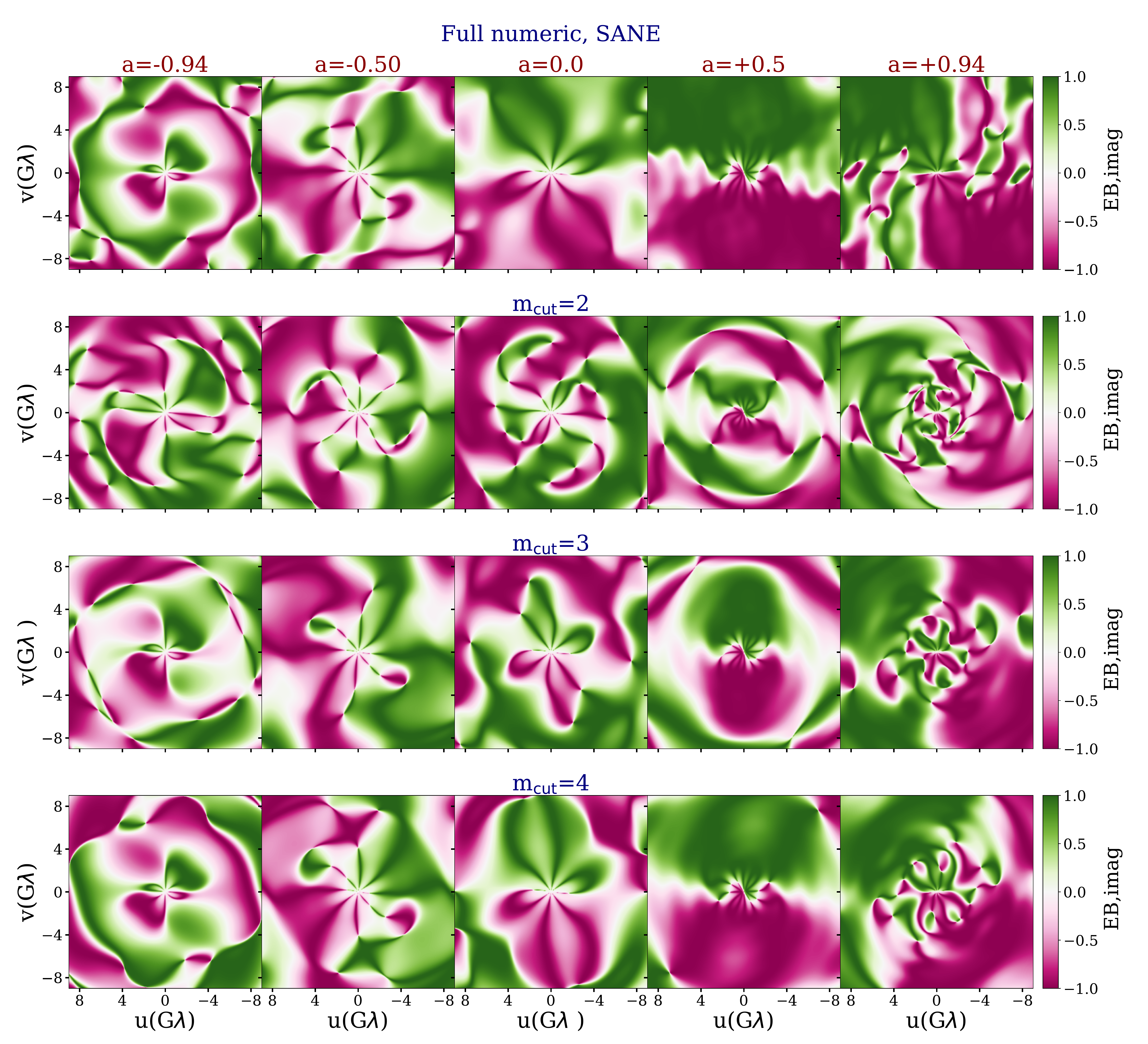}
\caption{The comparison between the imaginary part of the $EB$-correlation in SANE simulations from the full-numeric as well as with different $m$-order cuts. From the top to bottom, we present the full numeric, the case with  $m_{\mathrm{cut}}$ =2,  $m_{\mathrm{cut}}$ =3 and  $m_{\mathrm{cut}}$ =4, respectively. }
\label{Comparison_EB-Imagin_SANE}
\end{figure*}

Figures \ref{Comparison_EB-Real_MAD} and \ref{Comparison_EB-Real_SANE} present the real part of the $EB$-correlation function, while Figures \ref{Comparison_EB-Imagin_MAD} and \ref{Comparison_EB-Imagin_SANE} show the imaginary part of the $EB$-correlation function in MAD and SANE simulations, respectively. In each plot, from top to bottom, we present the full-numerical results, as inferred using the $M_1$ method in Section \ref{phase-1st-method}, as well as the results of the expansion using $m_{\mathrm{cut}} = 2$, $m_{\mathrm{cut}} = 3$ and $m_{\mathrm{cut}} = 4$, respectively. 

From the plots, it is evident that in all cases, there is a considerable contribution from modes beyond a truncation at $m_{\mathrm{cut}} = 2$.
Furthermore, it is seen that in MAD simulations, the real part is almost well represented up to $m_{\mathrm{cut}} = 3$. However, even in MAD simulations $m_{\mathrm{cut}} = 3$ is not completely sufficient for the imaginary part of the correlation function and we need to truncate it up to $m_{\mathrm{cut}} = 4$. The situation becomes more complicated for SANE simulations where neither the real nor the imaginary parts can be sufficiently described up to $m_{\mathrm{cut}} = 3$ and we need to consider up to at least $m_{\mathrm{cut}} = 4$. There are also some cases where higher order terms are required to fully describe the $EB$-correlation function, see for example $a = (+0.5, +0.94)$ in Figures \ref{Comparison_EB-Real_SANE} and \ref{Comparison_EB-Imagin_SANE}.

\begin{figure*}[th!]
\center
\includegraphics[width=0.99\textwidth]{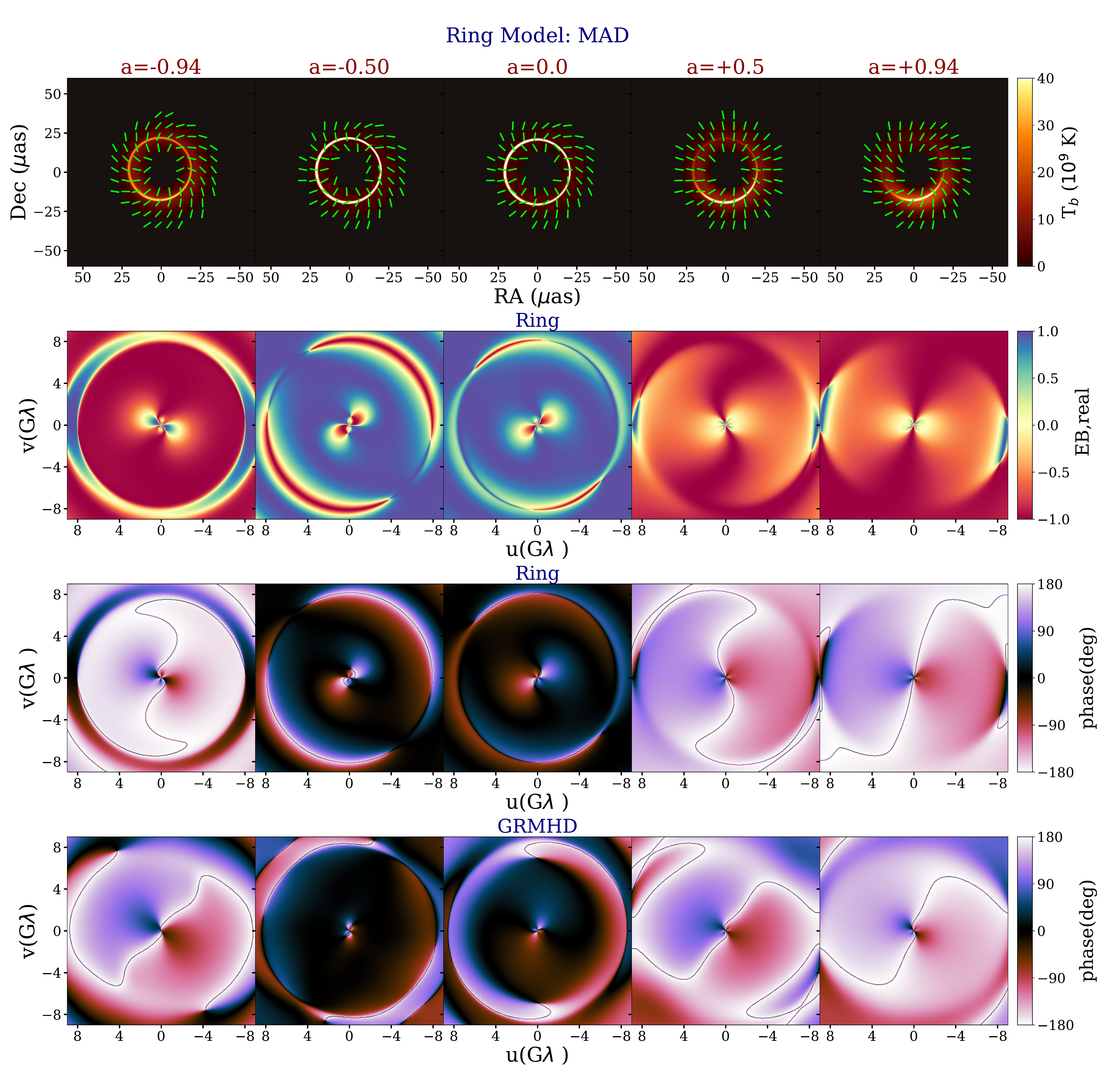}
\caption{The geometrical ring model vs. the time-averaged GRMHD MAD simulations. From the top to bottom we present the polarized images, the real part and the phases of the $EB$-correlation for the ring model and the phase from the time-averaged GRMHD simulation, respectively. It is inferred that the ring model recovers the full phase features in MADs. }
\label{Ring_MAD}
\end{figure*}

\begin{figure*}[th!]
\center
\includegraphics[width=0.99\textwidth]{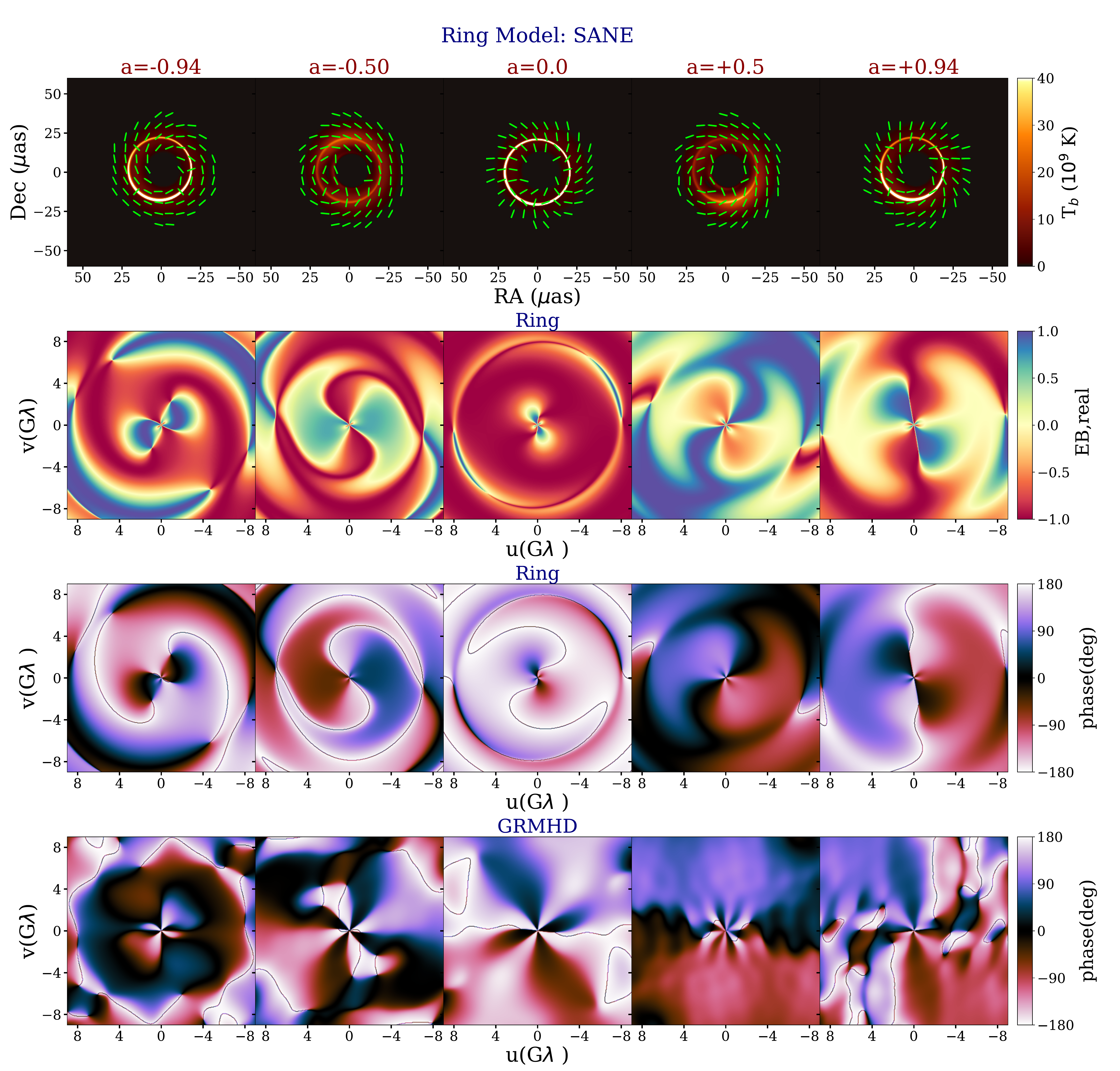}
\caption{The geometrical ring model vs. the time-averaged GRMHD SANE simulations. From the top to bottom we present the polarized images, the real part and the phases of the $EB$-correlation for the ring model and the phase from the time-averaged GRMHD simulation, respectively. It is inferred that the ring model does not seem to recover the full phase features in SANEs. }
\label{Ring_SANE}
\end{figure*}

\section{Constructing the geometrical Ring model}
\label{Geometrical-Ring}
Having presented the $EB$-correlation map for the synthetic data from the time-averaged GRMHD simulations of H-AMR, here we use a geometrical ring model, with free magnetic field geometry as well as the velocity field orientation in the equatorial plane. The aim is to vary the inclination of the magnetic field (hereafter as $i_B$ ), its orientation in the equatorial plane (specified with $\eta$) together with the velocity field orientation (identified with $\chi$) which are defined as:
\begin{equation}
\label{B-V-params}
\begin{split}
i_B  \equiv & \arctan{\left(B_z/\sqrt{B^2_x + B^2_y}\right)}, \\
\eta  \equiv & \arctan{\left(B_y/B_x \right)}, \\
\chi \equiv & \arctan{\left(V_y/V_x \right)}. \\
\end{split}
\end{equation}
Subsequently, we identify a subset of models that are well-matched with the time-averaged images and the $EB$-correlation functions in visibility space. Our analysis shows that out of a family of roughly 2000 different ring models, only about a few percent of them demonstrate great similarities with the GRMHD simulations. Table \ref{Table-Ring-model} presents the ring parameters associated with the consistent model with the time-averaged GRMHD simulations.

\begin{table*}[th!]
\hspace{-24pt}
\begin{tabular}{lccc|ccc|ccc|ccc|ccc}
\toprule
\hline
 & \multicolumn{3}{c}{a = -0.94} &  \multicolumn{3}{c}{a = -0.5} & \multicolumn{3}{c}{a = 0.0} & \multicolumn{3}{c}{a = +0.5} & \multicolumn{3}{c}{a = +0.94}  \\ \hline
Model & $i_B$ & $\eta$ & $\chi$ & $i_B$ & $\eta$ & $\chi$ & $i_B$ & $\eta$ & $\chi$ & $i_B$ & $\eta$ & $\chi$ & $i_B$ & $\eta$ & $\chi$  \\
\hline 
MAD & 59 & 0 & 108 & 59 & 0 & 144 & 59 & 0 & 144 & 59 & -72 & -144 & 59 & -108 & -108  \\ \hline 
SANE & 150 & 108 & 144 & 150 & 108 & 36 & 59 & 0 & -144 & 150 & -108 & -36 & 150 & 180 & -144 \\ \hline
\bottomrule 
\end{tabular}
\caption{The ring model parameters are similar to the time-averaged GRMHD simulations of MAD (top) and SANE (bottom). From left to right we increase the BH spin in $a = (-0.94, -0.5, 0.0, +0.5, +0.94)$. In each case, we present the magnetic field inclination, $i_B$, its orientation in the equatorial plane, $\eta$, together with the velocity field orientation, $\chi$, in the equatorial plane. }
\label{Table-Ring-model}
\end{table*}

Figures \ref{Ring_MAD} and \ref{Ring_SANE} present a ring-model approximation to the time-averaged GRMHD simulation for MAD and SANE cases, respectively. In each plot, from top-to-bottom, we present the ring image, the real part, and the phase of the $EB$-correlation from the ring model and the $EB$-correlation phase from the time-averaged H-AMR GRMHD simulations, respectively. In each row, from 
left to right, we increase the BH spin as $a = (-0.94, -0.5, 0.0, +0.5, +0.94)$. 

From the plots, it is inferred that the phase map in MAD simulations is almost fully recovered while it is qualitatively more distinct for SANE models, which is generally consistent with the  results of \cite{Emami2022}. 

\section{Comparison with the EHT 2017 data}
\label{Com-EHT-data}

As already stated above, this newly proposed $EB$-correlation function provides a complementary constraint to the previous analyses.
Using image-integrated net linear polarization $|m|_\mathrm{net}$, image-integrated net circular polarization $|v|_\mathrm{net}$, image-averaged linear polarization $\langle|m|\rangle$, and a coherent azimuthal structure measurement $\beta_2$ \citep[proposed by][]{2020ApJ...894..156P}, \citet{2021ApJ...910L..12E} shows strong evidence that M87* is in a MAD state. Specifically, SANE models tend to under produce $|m|_\mathrm{net}$ and $|\beta_2|$, and are disfavored. Therefore, we focus on MAD models here.

Having presented the EHT 2017 data for M87*, we make gridding in the visibility space using the EHT reconstructed image and we compare it with the time-averaged GRMHD simulations from H-AMR.

\begin{figure*}[th!]
\center
\includegraphics[width=0.99\textwidth]{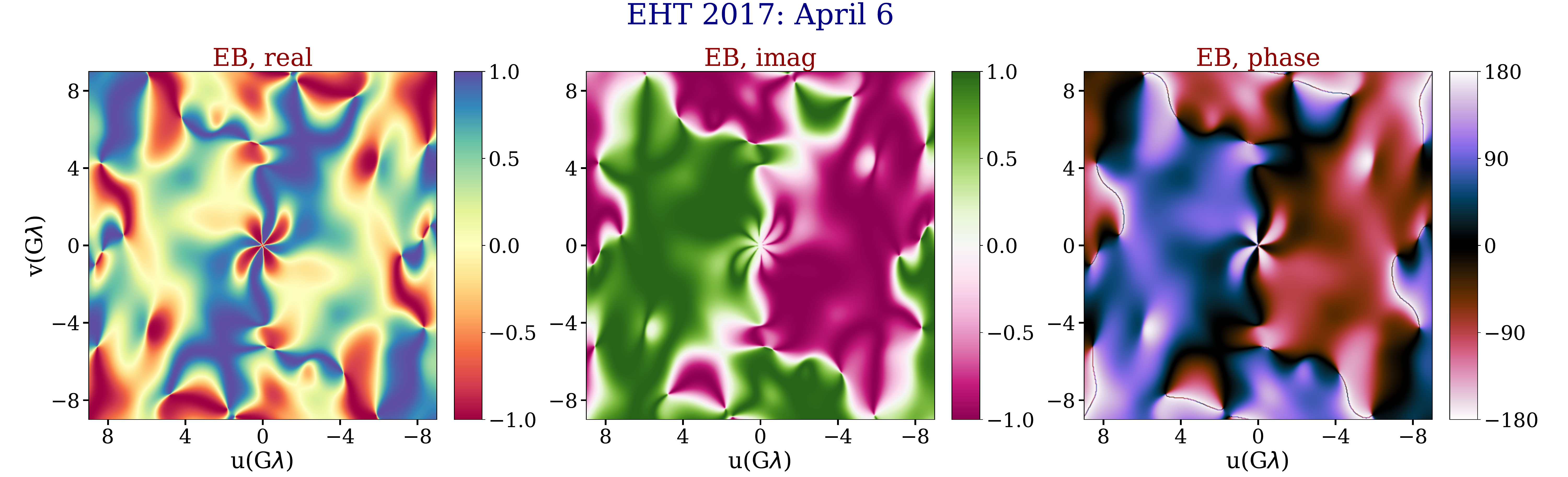}
\includegraphics[width=0.99\textwidth]{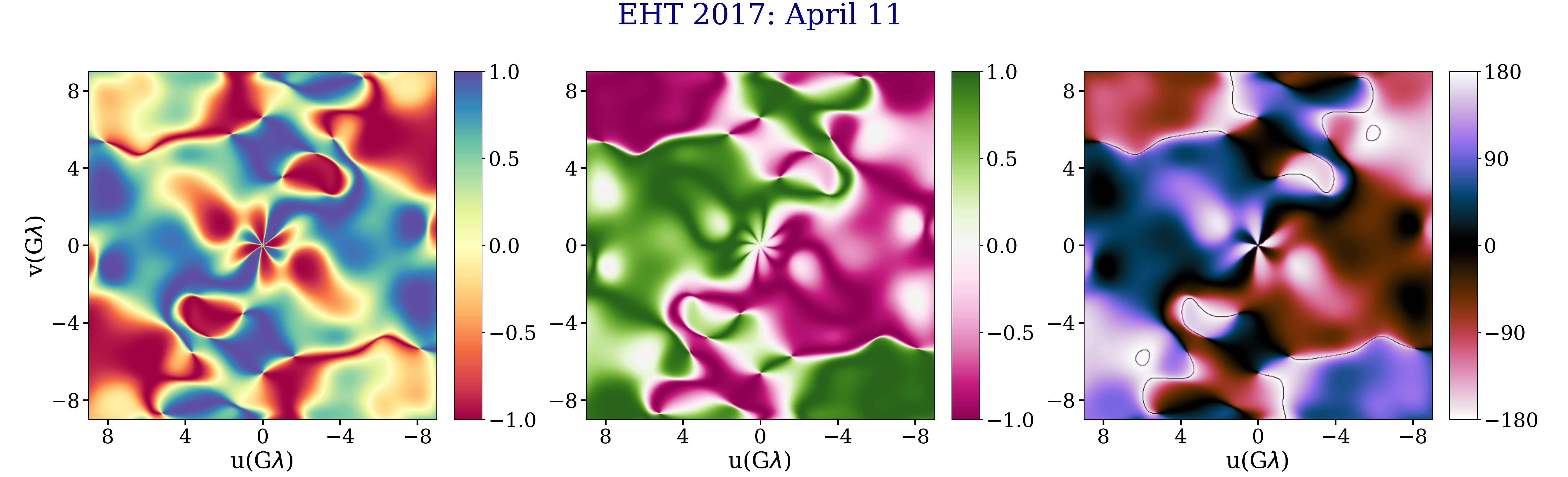}
\includegraphics[width=0.99\textwidth]{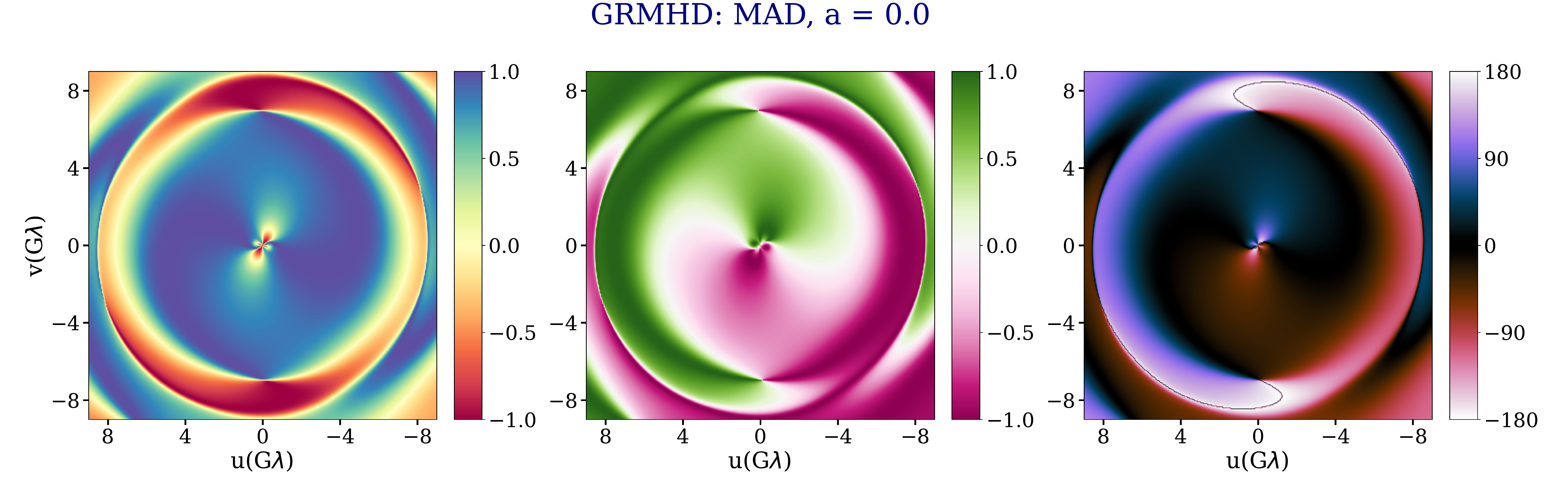}
\includegraphics[width=0.99\textwidth]{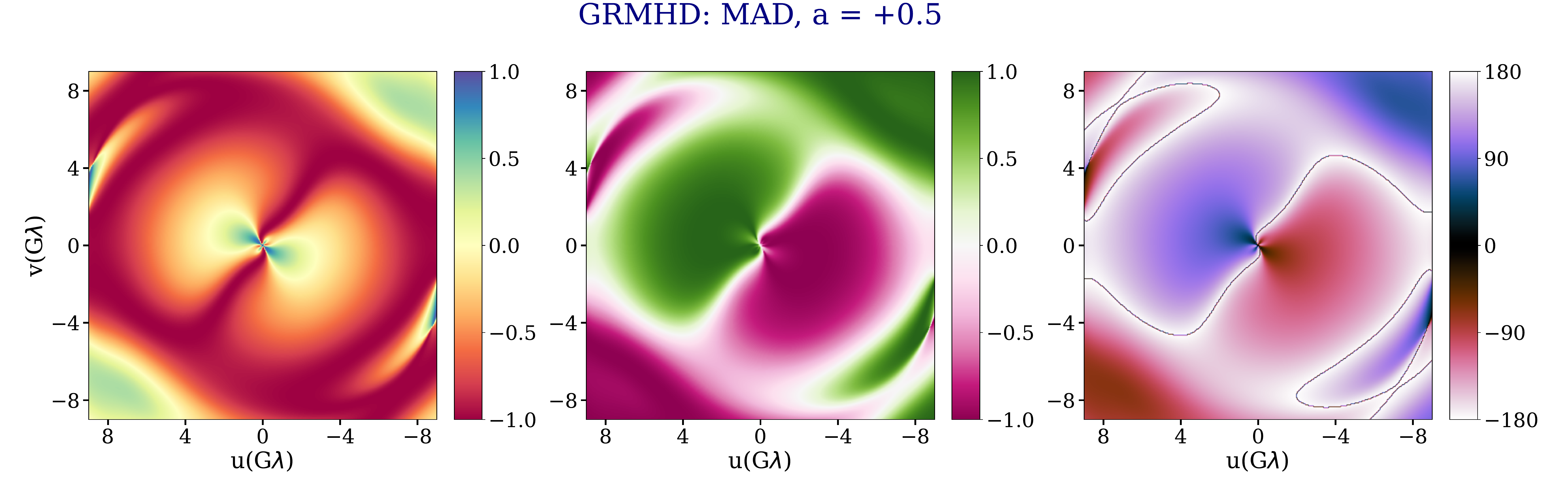}
\caption{The real, imaginary, and the phase of the gridded $EB$-correlation function for the EHT 2017 data for 3598 (1st-row) and 3601 (2nd-row) and the time-averaged GRMHD simulations in MAD with $a=0.0$ and $a=+0.5$ spins, respectively. The phase map in the EHT data contains both of the dark and bright regions which is similar to the MAD with $a=0.0$ in the dark regions while similar to MADs with $a=+0.5$ for bright regions. }
\label{EHT_data_grid}
\end{figure*}

In Figure \ref{EHT_data_grid} we present the real, imaginary, and the phase of the $EB$-correlation function for EHT 2017 on April 6 (1st-row), April 11 (2nd-row), and then from the time-averaged GRMHD simulations with $a = 0.0$ (3rd-row) and  $a=+0.5$ (4th-row), respectively. 

From the figure it is inferred that the EHT phase map contains both of the dark and bright regions, associated with low and intermediate phases. However the map of MAD case with $a=0$ mostly contains dark regions while the map from MAD with $a=+0.5$ merely contains bright regions. This demonstrates that either the BH spin sits somewhere in between and/or the electron temperature profile must be different in a favorite model. We leave this extra exploratory investigation to a future work with a goal to extend over the range of models and make a more direct comparison with the EHT data. 

\section{Conclusions}
\label{conclusion}
We introduced a new correlation function in the visibility space, the $EB$-correlation, and made an in-depth analysis of its real/imaginary parts as well as the phase using the EHT 2017 data for the M87* source together with the synthetic data from a subset of different toy models and the time-averaged GRMHD simulations from H-AMR. In the following, we review our key findings. 

$\bullet$ The phase map for the EHT 2017 data in April 6 and April 11 from Section \ref{EB-real-data} illustrates some patterns in the $EB$-correlation phase in Figure \ref{EHT-data} that includes both of dark and bright regions in the phase map. 

$\bullet$ It is inferred from a set of different toy models considered in Section \ref{Toy-Model-Set} that asymmetries in the EVPAs, as shown in Figure \ref{Gaussian_Model}, play important roles in launching non-zero phases. Furthermore, the phase structure from different geometrical ring models, shown in Figure \ref{G-Ring_Model} depends on different factors including the source inclinations together with the magnetic and the velocity field structures. 

$\bullet$ The phase map from different GRMHD simulations as studied in Section \ref{grid-based} present distinct features, as shown in Figures \ref{MAD-average-map-1st} and \ref{SANE-average-map-1st}, with some qualitative similarities in some cases that are predominant in MAD simulations. 

$\bullet$ Employing the Zernike polynomials done in Section \ref{ZP-expansion} as a new reconstruction algorithm for the $EB$-correlation phase map breaks degeneracies, shown in Figures \ref{Zernike_Moments_norder} and \ref{Zernike_Moments_morder}, between models that are otherwise qualitatively similar. 

$\bullet$ An azimuthal expansion of the $E$ and $B$ modes in terms of $m$-orders done in Section \ref{morder-expansion} demonstrates that 
to reconstruct both the real and the imaginary parts of the $EB$-correlation function we require considering higher order terms in the azimuthal expansion. More explicitly, in most cases considering the expansion up to $m_{\mathrm{cut}} = 4$, as shown in Figures \ref{Comparison_EB-Imagin_MAD} and \ref{Comparison_EB-Imagin_SANE}, recovers the majority of the features in the $EB$-correlation function map. 

$\bullet$ A set of geometrical ring models considered in Section \ref{Geometrical-Ring}, with distinct magnetic and velocity field geometries, shows that only a small set of models survive in recovering the patters in the $EB$-correlation function. Consequently, we argue that the $EB$-correlation function is capable of breaking the degeneracy between ring models that are otherwise very similar. 

$\bullet$ A qualitative comparison between the $EB$-correlation from the EHT 2017 data with the time-averaged GRMHD simulations, done in Section \ref{Com-EHT-data}, demonstrates that the phase map from the EHT data contains, as shown in Figure \ref{EHT_data_grid} a mixture of both of the dark and bright regions which partially appear in MAD simulations with $a=0.0$ (dark spots) while in MAD with $a=+0.5$ (bright spots). This suggests that to find out the favorite model we might need to change the BH spin as well as trying alternative electron temperature profiles. 

\section{Future direction}
\label{future}
While the current work focuses on a set of GRMHD simulations with 5 different BH spins with a limited set of electron temperature profile, in a future study we extend this analysis to also include more spins as well as  different electron temperature profiles. Furthermore, we aim to study cases with electron temperature being computed using 2 temperature fluid simulations \citep[e.g.,][]{2017MNRAS.466..705S}. The impact of the a tilted accretion disk would be also intriguing to be figured out.

\section*{Data Availability}
Data directly corresponding to this manuscript and the figures are available to be shared on reasonable request from the corresponding author. The ray tracing of the simulation done in this work was performed using the  {\sc ipole} method \citep{Moscibrodzka&Gammie2018}. We have used the library of H-AMR  simulations by \cite{2022ApJS..263...26L} from the standard library of 3D time-dependent GRMHD simulations in \citet{2022ApJ...930L..16E}. 

\section*{acknowledgement}
It is a great pleasure to thank  Geoffrey Bower, Andrew A. Chael, Michael Johnson, Daniel Palumbo, George Wong and the entire of the EHT collaboration for very fruitful conversations and comments. We also acknowledge the EHT internal review panel for constructive comments which improved the quality of this paper. Razieh Emami acknowledges the support from grant numbers 21-atp21-0077, NSF AST-1816420, and HST-GO-16173.001-A as well as the Institute for Theory and Computation at the Center for Astrophysics. Dominic Chang acknowledges the support of the Black Hole Initiative at Harvard University. We thank the supercomputer facility at Harvard University where most of the simulation work was done. This research was made possible through the support of grants from the Gordon and Betty Moore Foundation and the John Templeton Foundation. It was also supported by the National Science Foundation grants AST 1935980, AST 2034306, and OISE 1743747. The opinions expressed in this publication are those of the author(s) and do not necessarily reflect the views of the Moore or Templeton Foundations.

\textit{Software:} matplotlib \citep{2007CSE.....9...90H}, numpy \citep{2011CSE....13b..22V}, scipy \citep{2007CSE.....9c..10O}, seaborn \citep{2020zndo...3629446W}, pandas \citep{2021zndo...5203279R}, h5py \citep{2016arXiv160804904D}.

\appendix 
\section{Impact of the resolution}
\label{EHT-resolution}
\begin{figure*}[th!]
\center
\includegraphics[width=0.99\textwidth]{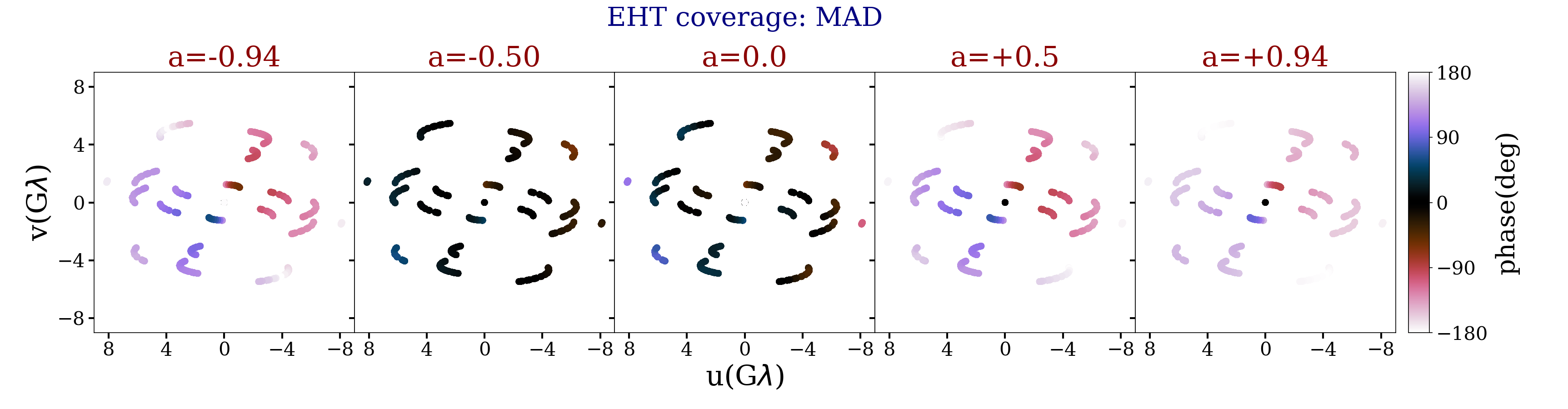}
\includegraphics[width=0.99\textwidth]{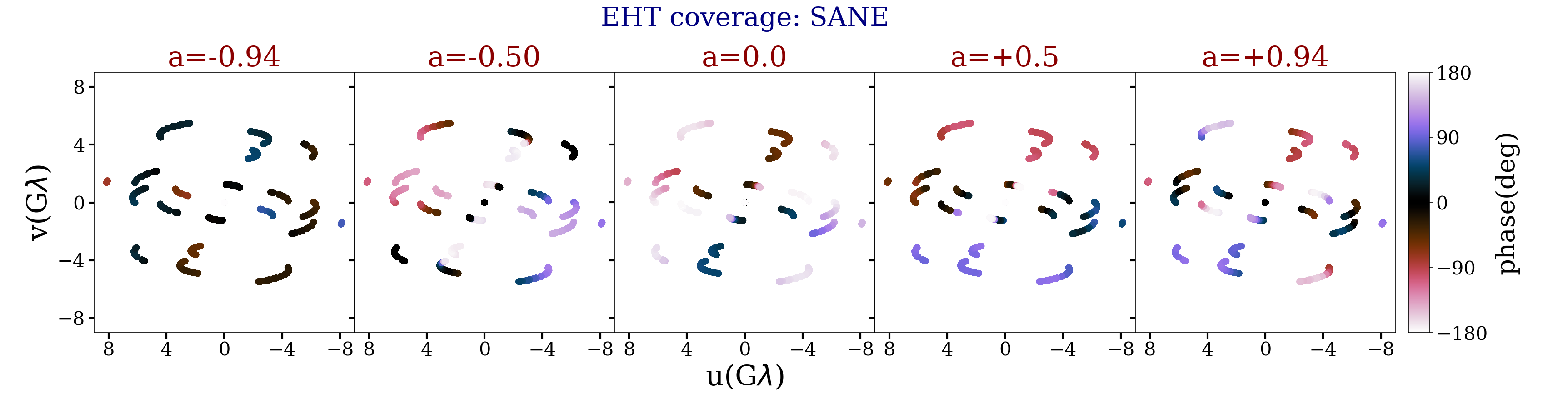}
\caption{The phase map for the $EB$-correlation for the time-averaged GRMHD simulations using the EHT coverage for M87* and taking into account the EHT finite resolution of 20 $\mu$as. Comparing this with Figure \ref{Phase-EHT-Coverage} demonstrates no noticeable differences. }
\label{Phase-EHT-Coverage_Blurred}
\end{figure*}

While the main text  solely focuses on the unblurred time-averaged GRMHD simulations, here we study the impact of adding the EHT finite resolution to the $EB$-correlation phase. Figure \ref{Phase-EHT-Coverage_Blurred} presents the time-averaged GRMHD simulation on top of the EHT coverage. The EHT finite resolution of 20 $\mu$as is taken into account in computing the time-averaging. Comparing this with the results from Figure \ref{Phase-EHT-Coverage}, it is seen that the phase map is very similar between these simulations as all what blurring do is to only shift the phase in the long baselines with no impacts for the short baselines. Consequently, the EHT resolution does not affect the $EB$-correlation phase.

\begin{figure*}[th!]
\center
\includegraphics[width=0.99\textwidth]{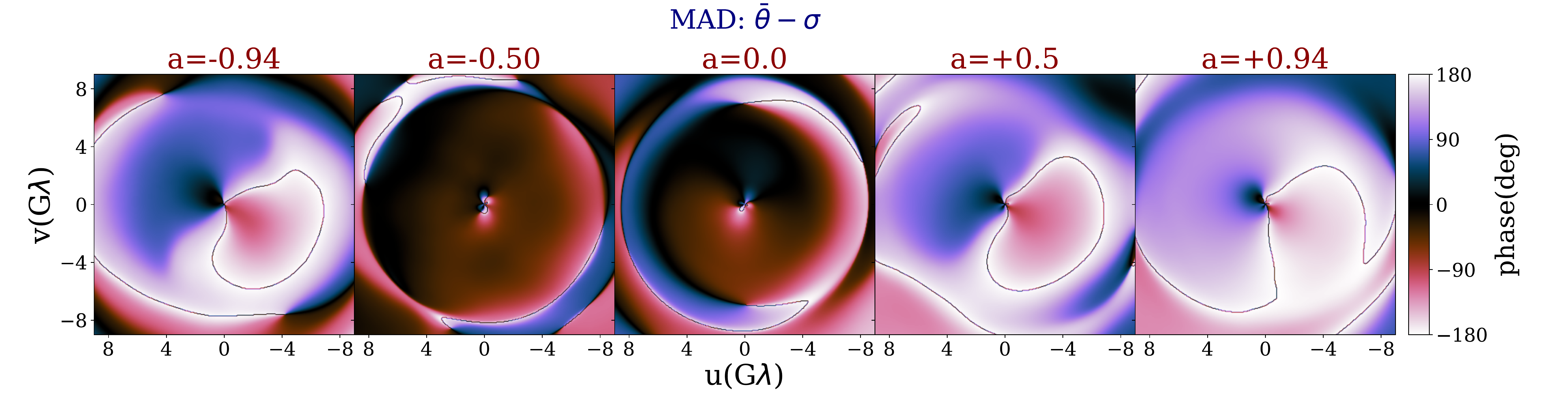}
\includegraphics[width=0.99\textwidth]{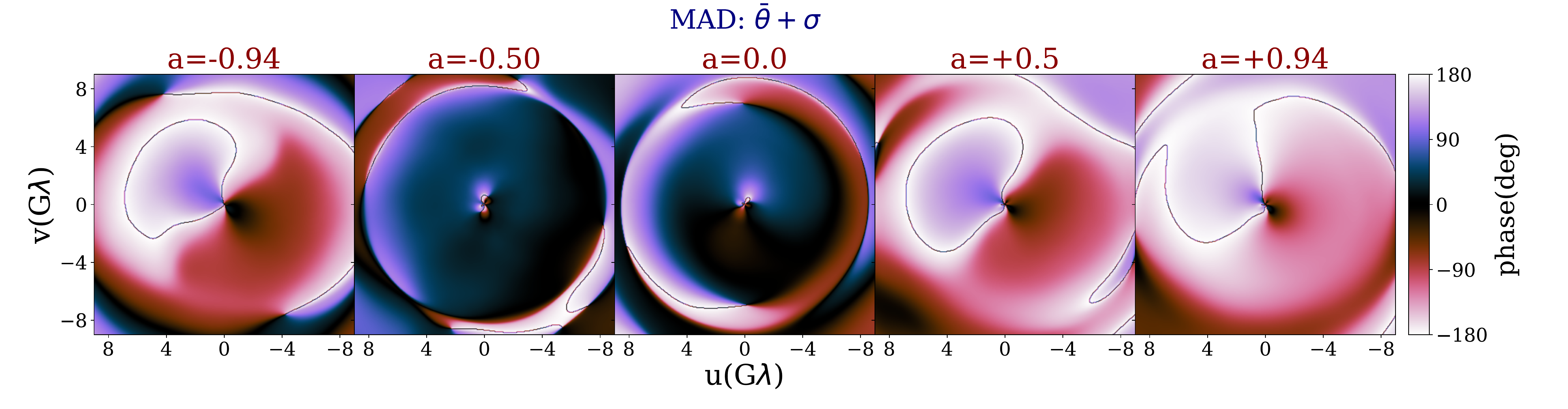}
\includegraphics[width=0.99\textwidth]{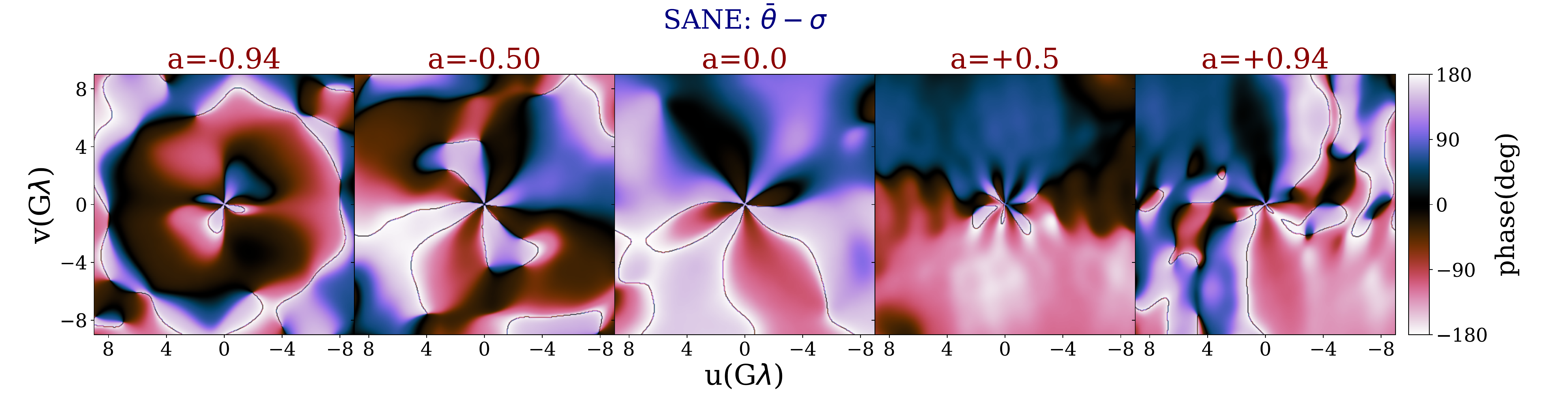}
\includegraphics[width=0.99\textwidth]{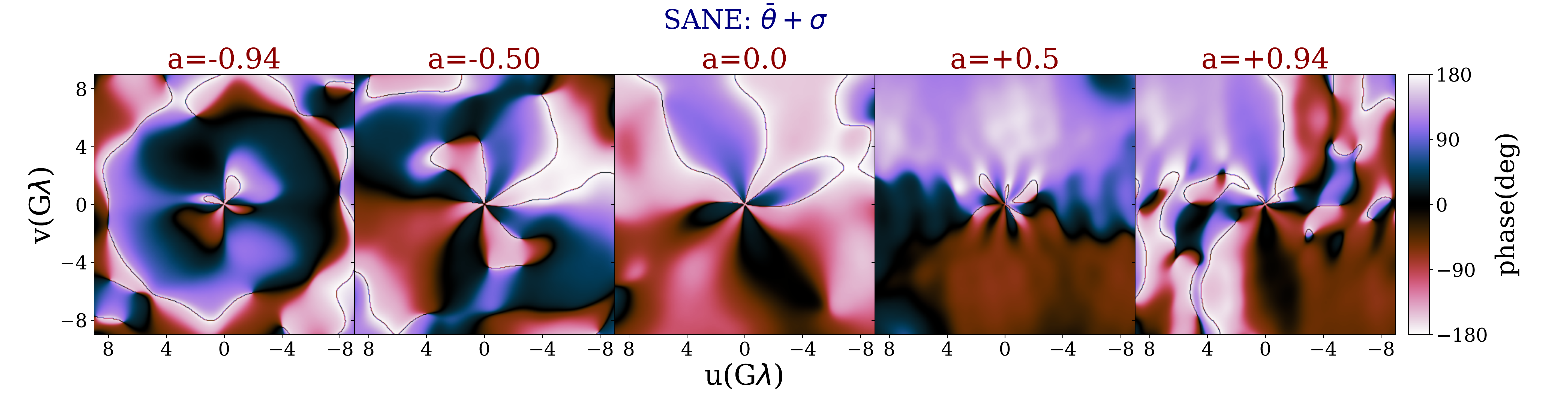}
\caption{The impact of time-variability as a variance in the phase map of the time-averaged $EB$-correlation map for MAD and SANE simulations with different BH spins. }
\label{Mad-SANE-average-variance}
\end{figure*}
\section{Impact of the time-variability in the mean phase }
\label{variance}

In the first order, the time variability in GRMHD simulations leads to an averaged change in the correlation phase map. To model this effect, we infer the phase variance and compute the phase map with this extra change being incorporated. Here we estimate the phase variance as: 
\begin{equation}
\label{variance-EB-phase}
\sigma (u,v) = f(a,b) \sqrt{ \left( \frac{\partial f(a,b)}{f \partial a} \Delta a \right)^2 +  \left( \frac{\partial f(a,b)}{f \partial b} \Delta b \right)^2},
\end{equation} 
where we have defined:
\begin{equation}
\label{f-im-re-def}
\begin{split}
f(a,b) \equiv &  \arctan{\left( \frac{ \langle Im(\rho_{\mathcal{EB}}(u,v)) \rangle}{ \langle Re(\rho_{\mathcal{EB}}(u,v)) \rangle } \right)} ,  \\
a \equiv & Im(\rho_{\mathcal{EB}}(u,v)), \\
b \equiv & Re(\rho_{\mathcal{EB}}(u,v)). 
\end{split}
\end{equation} 
More explicitly, we compute the phase map as:
\begin{equation}
\label{theta-modified}
\Theta = \bar{\theta}(u,v) \pm \sigma(u,v). 
\end{equation}
Figure \ref{Mad-SANE-average-variance} presents the $EB$-correlation phase map including the above variance. Descending rows refer to MADs and SANEs with phase variance being subtracted and added to the mean phase, respectively. 

From the plot, it is inferred that different simulations with various BH spins lead to substantially different phase maps with distinct features between the MAD and SANEs as well as different spins. Consequently, we argue that the phase of the $EB$-correlation could be a very good parameter to probe the BH spin as well as the accretion types near the BHs. Furthermore, the inclusion of the phase variance does not change the shape of the phase significantly. This means that at the leading order, we are not dominated by noise.

\section{Phase estimation from $\beta$ expansion} 
\label{phase-semi-analysis}
In this appendix, we walk through some of the steps in estimating the phase of the $EB$-correlation. Since the $E$ and $B$ modes are suppressed for higher values of $m$s, we truncate the expansion at $m_{\mathrm{cut}} = \pm 4$. The $E$ and $B$ modes can then be expressed as: 
\begin{align}
\label{E-B-expansion}
E = & E^r_{-4} + E^i_{-3} + E^r_{-2} + E^i_{-1} + E^r_0 + E^i_1 + E^r_2 + E^i_3 + E^r_4, \nonumber\\
B = & B^r_{-4} + B^i_{-3} + B^r_{-2} + B^i_{-1} + B^r_0 + B^i_1 + B^r_2 + B^i_3 + B^r_4.
\end{align}
where the super-index $r$ and $i$ refer to the real and imaginary parts of individual modes, respectively. While the sub-index numbers describe the $m$-order in our expansion.

From Eq. (\ref{E-B-expansion}) it is evident that the phase of the $EB$-correlation function will be non-zero only if we have non-equal $m$-modes in $E$ and $B$. Consequently, we split the $EB$-correlation into its real and imaginary components. Where the real part is computed as:
\begin{equation}
\begin{aligned}
\label{E-B-rr}
E(u,v) B^*(u,v) \bigg{|}_{\mathrm{real}} &=  \bigg{(} 
E^r_{-4} + E^r_{-2} + E^r_0 + E^r_2 + E^r_4 \bigg{)} 
\bigg{(} B^r_{-4} + B^r_{-2} + 
B^r_0 + B^r_2 + B^r_4 \bigg{)} - 
\bigg{(} E^i_{-3} +  E^i_{-1} + E^i_1 + E^i_3 \bigg{)} \bigg{(}  B^i_{-3} + B^i_{-1} + \\
& B^i_1 + B^i_3 \bigg{)},
\end{aligned}
\end{equation}
while the imaginary component is computed as: 
\begin{equation}
\begin{aligned}
\label{E-B-im}
E(u,v) B^*(u,v) \bigg{|}_{\mathrm{imag}} &= - \bigg{(} 
E^r_{-4} + E^r_{-2} + E^r_0 + E^r_2 + E^r_4 \bigg{)}  \bigg{(}  B^i_{-3} +
 B^i_{-1} + B^i_1 + B^i_3 \bigg{)} + \bigg{(} 
E^i_{-3} + E^i_{-1} + E^i_1 + E^i_3 \bigg{)} 
\bigg{(} 
B^r_{-4} + B^r_{-2} + \\
& B^r_0 +  B^r_2 + B^r_4 \bigg{)} .
\end{aligned}
\end{equation}
As already stated above, the $EB$-correlation function mixes the even and odd components of the $E$ and $B$ modes. As is explained in the main text, it requires us to include higher order terms in the $E$ and $B$ mode expansion. 

\bibliography{main}
\end{document}